\documentclass{aa}  

\usepackage{graphicx}
\usepackage{txfonts}
\usepackage{todonotes}
\begin{document}

   \title{Characterisation of the stellar wind in Cyg~X-1 via modelling of colour-colour diagrams}

   \author{
   E.~V.~Lai \inst{\ref{affil:INAF},\ref{affil:CAMK}} \and B.~De~Marco\inst{\ref{affil:EEBE}} \and
   Y.~Cavecchi \inst{\ref{affil:EEBE}} \and
   I.~El~Mellah \inst{\ref{affil:Univ_Santiago},\ref{affil:USACH}} \and
   M.~Cinus \inst{\ref{affil:UMK}} \and
   C.~M.~Diez \inst{\ref{affil:ESAC}} \and
   V.~Grinberg \inst{\ref{affil:ESTEC}} \and 
   A.~A.~Zdziarski \inst{\ref{affil:CAMK}} \and
   P.~Uttley \inst{\ref{affil:Pannekoek}} \and
   M.~Bachetti \inst{\ref{affil:INAF}} \and
   J.~José \inst{\ref{affil:EEBE}} \and
   G.~Sala \inst{\ref{affil:EEBE},\ref{affil:IEEC}}\and
   A.~Różańska \inst{\ref{affil:CAMK}} \and
   J.~Wilms \inst{\ref{affil:Remeis}} }
   
\institute{
INAF-Osservatorio Astronomico di Cagliari, via della Scienza 5, I-09047, Selargius (CA), Italy \label{affil:INAF}, \email{eleonora.lai@inaf.it}
\and
Nicolaus Copernicus Astronomical Center, Polish Academy of Sciences, Bartycka 18, PL-00-716 Warszawa \label{affil:CAMK} 
\and
Departament de Física, EEBE, Universitat Politècnica de Catalunya, c/Eduard Maristany 16, 08019 Barcelona, Spain  \label{affil:EEBE} 
\and
Departamento de Física, Universidad de Santiago de Chile, Av. Victor Jara 3659, Santiago, Chile \label{affil:Univ_Santiago} 
\and
Center for Interdisciplinary Research in Astrophysics and Space Exploration (CIRAS), USACH, Chile \label{affil:USACH} 
\and
Institute of Astronomy, Faculty of Physics, Astronomy and Informatics, Nicolaus Copernicus University, Grudziądzka 5, 87-100 Toruń, Poland \label{affil:UMK} 
\and
European Space Agency (ESA), European Space Astronomy Centre (ESAC), Camino Bajo del Castillo s/n, 28692 Villanueva de la Cañada, Madrid, Spain \label{affil:ESAC}
\and European Space Agency (ESA), European Space Research and Technology Centre (ESTEC), Keplerlaan 1, 2201 AZ Noordwijk, The Netherlands \label{affil:ESTEC}
\and
Anton Pannekoek Institute, University of Amsterdam, Science Park 904, 1098 XH Amsterdam, The Netherlands \label{affil:Pannekoek}
\and
Institut d’Estudis Espacials de Catalunya, c/Gran Capità 2-4, Ed. Nexus-201, 08034 Barcelona, Spain \label{affil:IEEC}
\and 
Dr. Karl Remeis-Observatory, University of Erlangen-Nuremberg, Sternwartstr. 7, 96049 Bamberg, Germany \label{affil:Remeis}
}

 
  \abstract
   {Cygnus X-1 is a high mass X-ray binary where accretion onto the black hole is mediated by the stellar wind from the blue supergiant companion star HDE~226868. Due to its inclination, the system is a perfect laboratory to study the not yet well-understood stellar wind structure. In fact, depending on the position of the black hole along the orbit, X-ray observations can probe different layers of the stellar wind. Deeper wind layers can be investigated at superior conjunction (i.e. null orbital phases).} 
   {We aim at characterising the stellar wind in the Cyg~X-1/HDE~226868 system analysing one passage at superior conjunction covered by {\it XMM-Newton} during the CHOCBOX campaign.}
   {To analyze the properties of the stellar wind we computed colour-colour diagrams. Since X-ray absorption is energy-dependent, colour indices provide information on the parameters of the stellar wind, such as the column density $N_\mathrm{H,w}$ and the covering factor $f_c$. We fitted  colour-colour diagrams with models that include both a continuum and a stellar wind component. We used the kernel density estimation (KDE) method to infer the unknown probability distribution of the data points in the colour-colour diagram, and selected the model corresponding to the highest likelihood. In order to study the temporal evolution of the wind around superior conjunction, we extracted and fitted time-resolved colour-colour diagrams.} 
   {We found that the model that best describes the shape of the colour-colour diagram of Cyg~X-1 at superior conjunction requires the wind to be partially ionised.
   The shape of the colour-colour diagram strongly varies during the analysed observation, as due to concurrent changes of the mean $N_\mathrm{H,w}$ 
   and the $f_c$ of the wind. Our results suggest the existence of a linear scaling between the rapid variability amplitude of $N_{\rm{H,w}}$ (on time scales between 10 s and 11 ks) and its long term variations (on time scales $>11$ ks). Using the inferred best-fit values, we estimated the stellar mass loss rate to be $\sim 7\times10^{-6} {\rm M_{\odot}yr^{-1}}$ and the clumps to have a characteristic mass of $\sim10^{17}\ {\rm g}$.}   
   {}
   

   \maketitle

%

\section{Introduction}

High mass X-ray binaries (HMXBs) are a class of binary systems in which a neutron star (NS) or a black hole (BH) is fed by the stellar wind from a massive ($\gtrsim 10\,M_{\odot}$) OB companion or by the decretion disc surrounding a quickly spinning Be donor star. The stellar wind is line-driven \citep{Castor_1975} and can be accelerated up to terminal velocities of $v_{\infty}\sim2500$~km~$\rm{s}^{-1}$ \citep[e.g.][]{Nunez_2017}. Internal shocks lead to the formation of overdense regions, or clumps, making the wind inhomogeneous and highly structured \citep[e.g.][]{Owocki_1988,Feldmeier_1997,Oskinova_2012,Sundqvist_Owocki_2013}.
While several stellar wind models have been developed \citep[e.g.][]{Oskinova_2012,Sundqvist_Owocki_2013,Sundqvist_2018,ElMellah_2018,ElMellah_2020}, the physical properties of the clumps, such as their shape, density and ionisation structure, are yet to be precisely constrained.  
Investigating the structure of the stellar winds produced by massive stars is crucial for various purposes, including constraining models of stellar and galactic evolution, and understanding the accretion mechanism in wind-fed HMXBs.\\
As a matter of fact, stellar winds represent an important mechanism of stellar mass loss, with mass-loss rates up to $10^{-5}\ \rm{M_{\odot}\,yr^{-1}}$ \citep{Puls2008,Nunez_2017}. This has an influence on the properties of the donor star (such as mass, luminosity, spin, chemical composition) as well as its evolutionary timescales. Moreover, in binary systems interacting via mass transfer mediated by a stellar wind like most HMXBs, only a fraction of the expelled gas is accreted, while the rest escapes the system, causing a removal of angular momentum and a change in the orbital parameters. Whether this process leads to a shrinking or a widening of the orbit \citep[e.g.][]{Hoyle1939,Bondi1944,Paczynski1976,ElMellah_2020} depends on key parameters, such as the wind terminal velocity, the initial orbital separation, and the accretion efficiency \citep[e.g.][]{Saladino2018}. The evolutionary pathways of such interactions can have significant impact on the rate of merging events and guide us in the quest for merger progenitors \citep[e.g.][]{Bulik_2011,Belczynski_2011,Neijssel_2021}.

The well-known HMXB system Cygnus X-1 (Cyg~X-1 hereafter) is a good target to constrain the stellar wind structure and infer information on the physical properties of wind clumps. 
The system hosts a BH with $M_\mathrm{BH}=21.2 \pm 2.2\, \rm{M_{\odot}}$ \citep{MillerJones_2021} in a $\sim 5.6\,\rm{d}$ quasi-circular orbit \citep{Gies_2003}. The $\rm{O}9.7\,\rm{Iab}$ supergiant companion star HDE~226868, with $M_{\ast}=40.6^{+7.7}_{-7.1}\,\rm{M_{\odot}}$ and $R_{\ast}=22.3\pm1.8\,\rm{R_{\odot}}$ \citep{MillerJones_2021}, launches fast stellar winds \citep[$v_{\infty}=2100\ {\rm km\,s^{-1}}$,][]{Herrero_1995}. The orbital inclination of the system \citep[$i\sim27^{\circ}$,][]{MillerJones_2021} causes our line of sight (LOS) to intercept the clump forming region, likely located close to the companion star's surface \citep[e.g.][]{ElMellah_2020}. \\
In this system, the presence of the wind significantly modifies the spectral \citep[e.g.][]{Nowak_2011} and timing \citep[][]{Lai_2022} properties of the X-ray source. Indeed, X-ray photons undergo different levels of absorption depending on the amount and the ionisation state of the wind material intercepting the LOS. 
Wind absorption mostly affects the softest energy bands ($E\lesssim1$ keV), although its influence can be observed up to higher energies $E\sim8$--$10$ keV \citep[e.g.][]{Hirsch_2019,Grinberg_2015}. The strength of the absorption events strongly varies as a function of the orbital phase \citep[e.g.][]{Balucinska_2000,Poutanen_2008,Hirsch_2019,Grinberg_2020,Lai_2022}. In particular, the X-ray light curves show more intense and frequent absorption dips when the X-ray source is at superior conjunction ($\phi_{\mathrm{orb}}=0$), i.e. when the BH is the farthest and the companion star the closest from the observer along the orbit 
(see \citeauthor{Li_Clark_1974} \citeyear{Li_Clark_1974} and references therein). In \cite{Lai_2022}, we showed that the absorption dips significantly contribute to the X-ray variability of Cyg~X-1 by increasing its observed fractional variability on timescales longer than $\sim 1$ s, and suggested the motion of clumps of the size of $\sim 10^{-4}\,R_{\ast}$ to be responsible for the enhancement of X-ray variability on these timescales.

Radiative hydrodynamical simulations show that the size of the clumps should increase as they move away from the star \citep[][]{Sundqvist_2018}, thus producing dips of increasingly longer duration. Therefore, to fully investigate the properties of the clumps, we need to access a broad range of timescales, down to the shortest timescales ($\sim 1$ min or shorter) characterising the smallest clumps formed close to the base of the wind.
However, the limitations of current instruments 
preclude wind-induced spectral changes occurring on very short timescales from being investigated via canonical spectral analysis. 

An alternative, powerful approach is to use colour-colour diagrams \citep[e.g.][]{Nowak_2011,Grinberg_2020}. 
A detailed modelling of the colour-colour diagram of wind-fed systems can give constraints on the physical properties of the wind. Earlier studies made use of empirical functions or simple (neutral) stellar wind models to describe the characteristic shape of colour-colour tracks \citep[e.g.][]{Hanke_2008,Nowak_2011,Hirsch_2019}.
Recently, \citet{Grinberg_2020} explored more complex wind models, concluding that a proper description of the observed colour-colour tracks requires accounting for different levels of ionisation in the wind.
In this paper, building on the work of \citet{Grinberg_2020}, we model the colour-colour diagrams of Cyg~X-1 in order to constrain the parameters of the wind as a function of the orbital phase. This information can be used to infer physical properties such as the stellar mass-loss rate and the mass of the clumps \citep[e.g.][]{ElMellah_2020}. 

In this paper, we first describe the data reduction procedure (Sect.~\ref{reduction}), then we test a number of 
stellar wind models (Sect.~\ref{Colour-colour diagrams}) to determine the one that best describes the colour-colour diagram of Cyg~X-1 in its hard state (the most affected by the wind, \citealt{Nowak_2011}). 
In Sect.~\ref{time-resolved}, we investigate the temporal evolution of the column density and covering factor of the wind by fitting time-resolved colour-colour diagrams. We discuss our results in Sect.~\ref{discussion} and present our conclusions in Sect.~\ref{conclusion}.

\section{Observation and data reduction}\label{reduction}
\begin{figure}
\centering
\includegraphics[width=1\linewidth]{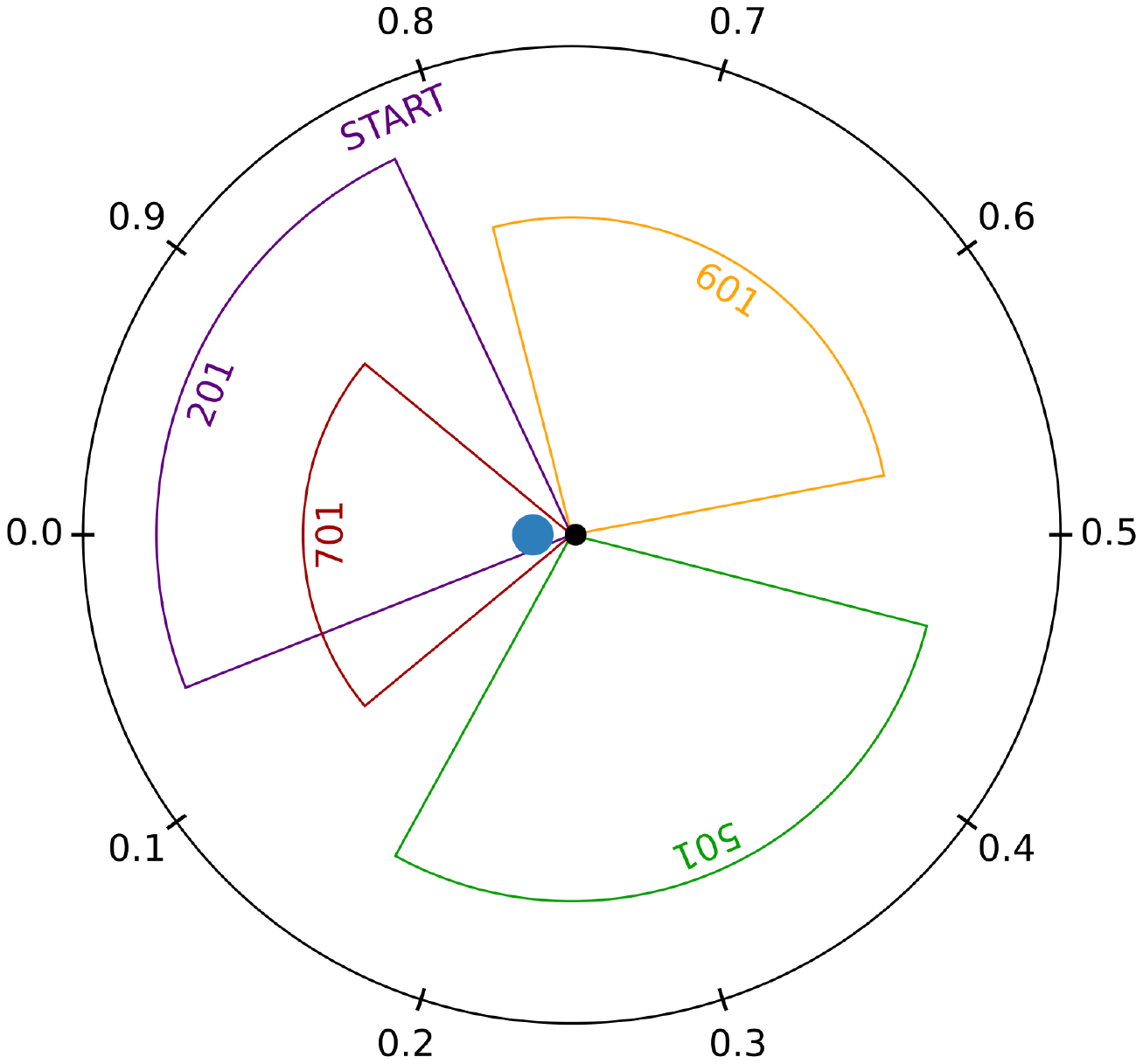}
    \caption{Orbital phase coverage of the {\it XMM-Newton} CHOCBOX monitoring of Cyg~X-1 in the hard state. The four arc labels correspond to the ObsIDs of each pointing (see table 1 in \citealt{Lai_2022} for the full ObsID). The orbital phase $\phi_{\mathrm{orb}}=0$ indicates the passage at superior conjunction. Observation 201 (in purple) is the focus of the analyses presented in this paper. The starting orbital phase of the monitoring (corresponding to $\phi_{\mathrm{orb}}=0.82$) is also indicated. HDE~226868 and Cyg~X-1 are only schematically represented, no Roche lobe was graphically considered in the scheme.}
    \label{fig:chocbox}
\end{figure}

We focused our analysis on a single {\it XMM-Newton} observation of Cyg~X-1 (obsID 0745250201, hereafter 201) in its hard state, which is part of the $\sim 1.5$ orbital period-long monitoring from the multi wavelength campaign ``Cyg~X-1 Hard state Observations of a Complete Binary Orbit in X-rays" (CHOCBOX). The full CHOCBOX coverage is shown in Fig.~\ref{fig:chocbox}. The monitoring consists of four pointings for a total duration of $\sim 7$ days, and catches two consecutive passages of the X-ray source at superior conjunction. \\
In \cite{Lai_2022} we investigated the changes driven by the stellar wind in the X-ray spectral-timing properties of the source throughout the entire monitoring. Among all the CHOCBOX observations, observation 201 turned out to be the most affected by wind absorption. It also showed the most prominent effects on the X-ray variability on short timescales (in the $\sim 0.1$--$10$ s range) as compared to the other observations. 
Observation 201 covers orbital phases $\phi_{\mathrm{orb}} = 0.82$--$0.06$, corresponding to the first passage at superior conjunction \citep[][]{Lai_2022}. 
We focused on observation 201 for the study of the stellar wind, although in Sect.~\ref{time-resolved} we also used observation ID 0745250501 (hereafter 501) to identify the region of the colour-colour diagram less affected by wind-absorption. Indeed, observation 501 covers the orbital phases $\phi_{\mathrm{orb}} = 0.17$--$0.46$. Being the closest to inferior conjunction with $\phi_{\mathrm{orb}} = 0.5$ it is also the least absorbed. 
For both observations we used EPIC-pn data \citep[][]{Struder2001} in timing observing mode.
We performed the data reduction using the {\it XMM-Newton} \texttt{Science  Analysis Software (SAS}, version 20.0.0) and calibration files (CCF) as of 2022 March.

\cite{Diez_2023} showed that the extraction of EPIC-pn spectra in timing mode using the default Rate Dependent PHA (RDPHA) correction led to energies higher than expected for line features, and suggest a non-standard calibration to mitigate the problem.
We followed the same approach proposed by \cite{Diez_2023}. We ran \texttt{epproc} on the Observation Data Files (ODFs), turning off the RDPHA correction (\texttt{withrdpha=’N’}) and applying the Rate Dependent CTI (RDCTI) correction using \texttt{epfast} (\texttt{runepfast=’Y’}).
Calibrated event files were screened for the presence of strong background flaring events using light curves with 1s time resolution in the $10$--$15$ keV energy band. No significant flares were observed. 
Using the \texttt{SAS} task \texttt{epatplot}, we found that a small fraction of pile-up is present in the data. We corrected for it by extracting source counts from a region which excludes the central pixels (RAWX $ = [36:39]$), i.e. RAWX $ = [30:35;\ 40:46]$. 
Background counts cannot be reliably estimated from timing mode observations of bright sources, as the entire CCD is illuminated by source photons. 
Since the PSF is energy dependent, extracting the background from the outer columns is not recommended as this procedure modifies the intrinsic spectrum \citep{Ng_2010}.
Since the source is very bright, with $\gtrsim 220$ counts s$^{-1}$ in the $0.5$--$10$ keV energy band, the background contribution is expected to be negligible. 
Therefore, we decided to not account for the background in our analysis.
We made use of the \texttt{SAS} tasks \texttt{arfgen} and \texttt{rmfgen} to extract the Ancillary Response Files (ARF) and Redistribution Matrix Files (RMF). We produced two ARF files: one for the full region (RAWX $= [30:46]$) and one for the central excluded region (RAWX $= [36:39]$). Using the command \texttt{addarf}, we subtracted the latter from the former, generating the final ARF file \citep[e.g.][]{Wilkinson_Uttley2009, Lai_2022}. The spectrum was rebinned to ensure a minimum of 20 counts per bin in order to apply chi-square statistics. \\
Spectral fits were done using \texttt{XSPEC} v12.10.1 \citep{Arnaud_1996}. We used the Interactive Spectral Interpretation System (\texttt{ISIS}) version 1.6.2 and a custom code, written in Python 3.10, for the calculation and the modelling of the colour-colour diagrams. Time series were extracted using \texttt{stingray} v1.1.2 \citep{Huppenkothen_2019, Huppenkothen2019, Bachetti_2023}


\section{Models for the colour-colour diagram of Cyg~X-1}\label{Colour-colour diagrams}

\begin{figure*}
\centering
\includegraphics[width=1\linewidth]{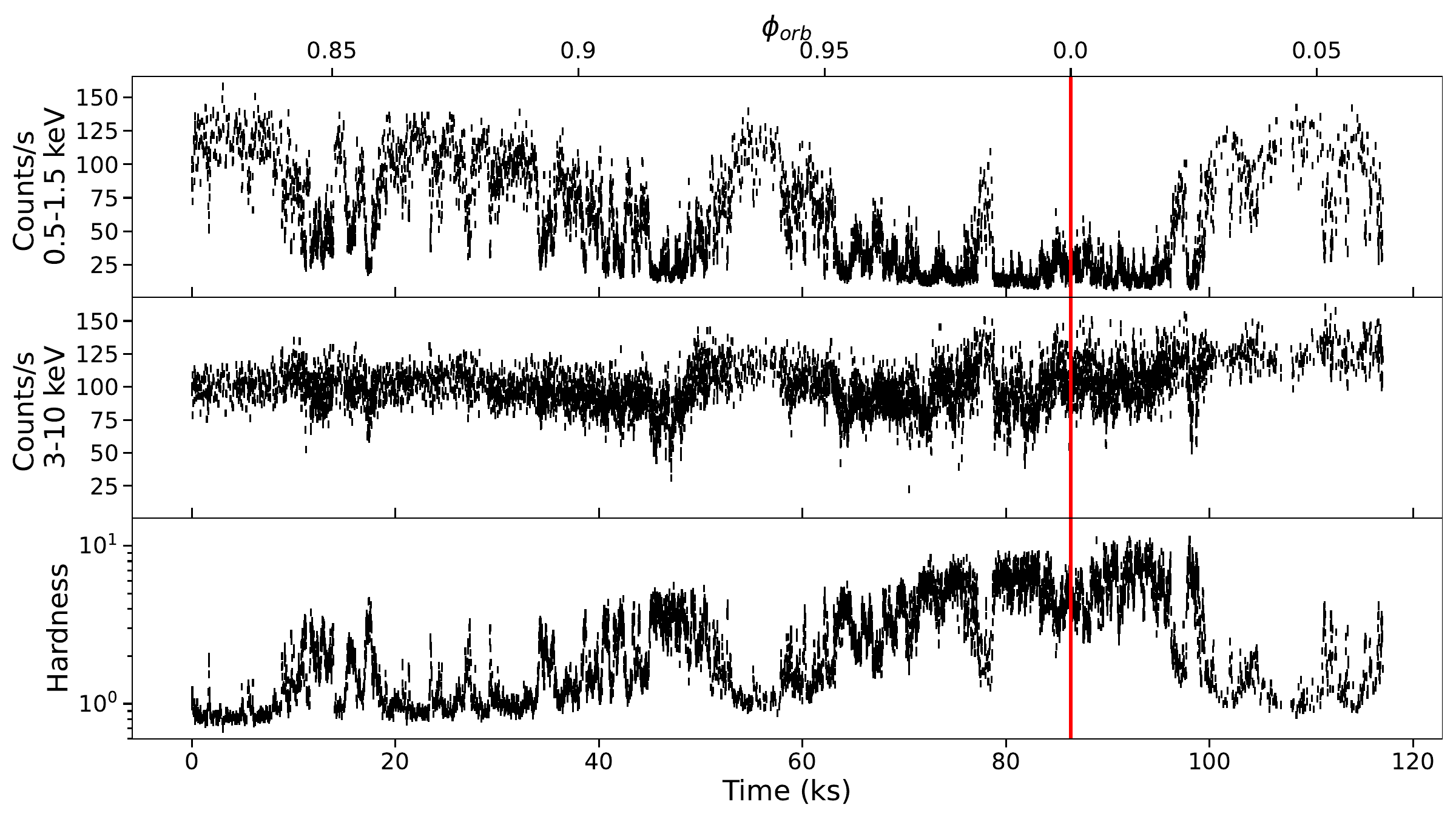}
    \caption{{\it XMM–Newton} EPIC-pn light curves of the observation 201 of Cyg~X-1 with time bins of 10 s. The upper and the middle panel show, respectively, the $0.5$--$1.5\,\rm{keV}$ and the $3$--$10\,\rm{keV}$ light curves. The bottom panel reports the hardness, i.e. ratio between count rates in the $3$--$10\,\rm{keV}$ and $0.5$--$1.5\,\rm{keV}$ energy bands. The red vertical line indicates the passage at superior conjunction (i.e. $\phi_{\mathrm{orb}}=0$).}
    \label{lc_hardness_201}
\end{figure*}
The passage of absorbing clumps across the LOS manifests as dips in the X-ray light curve of length ranging between several seconds to hours \citep[e.g.][]{Hirsch_2019}. To explore the presence and duration of dips in observation 201, we show its light curve in Fig.~\ref{lc_hardness_201}. In the upper and middle panels, we report the soft ($0.5$--$1.5\,\rm{keV}$) and the hard ($3$--$10\,\rm{keV}$) band light curves, respectively.
For the light curves of the entire monitoring, see \cite{Lai_2022}\footnote{Note that the correct starting time of the observation is 57535.9~MJD, instead of the 57535.0~MJD date reported in the caption of figure 1 of \cite{Lai_2022}.}. \\
The presence of dips in the soft band light curve, that are approximately three times less intense in the hard band light curve, produces sharp increases of the hardness ratios (as shown in the bottom panel of Fig.~\ref{lc_hardness_201}, where we choose to compare two non-contiguous energy bands to maximise the intensity of the dips). 
The single dips are quite short ($\lesssim 1$ ks). Therefore, given the swiftness of such events, their energy spectrum is difficult to study via time-resolved spectroscopy.
An alternative approach is to investigate the broad band spectral variability caused by wind clumps using colour-colour diagrams \citep[e.g.][]{Nowak_2011}.
These diagrams map the evolution of the ratio of the count rates in different energy bands. Specifically, a soft colour, defined as the ratio between count rates in a soft and an intermediate energy band, is plotted against a hard colour, defined as the ratio between count rates in the same intermediate band and a hard energy band. 
High values of both soft and hard colours characterise the least absorbed phases, thus mapping the upper region of the diagram. As absorption along the LOS increases, the hard and soft colours change. In particular, for a partially covering absorber the resulting colour-colour tracks describe a "pointy" or "nose-like" shape as the source becomes more absorbed \citep[i.e. in the lower region of the colour-colour diagram, see e.g. ][]{Hirsch_2019}. 
Following previous studies  \citep{Hanke_2008,Nowak_2011,Hirsch_2019,Grinberg_2020}, we constructed the colour-colour diagram of observation 201 using light curve bins of 10\,s, thus allowing us to detect also the spectral variability that could be caused by small clumps. The size of the clumps is directly related to the fly-by time across the LOS \citep{ElMellah_2020}. With this time resolution, the minimum size of the clumps that can be tested is $\sim2\times10^{-4} R_{\ast}$, assuming a wind terminal velocity of $v_{\infty}=2400\,\mathrm{km}\, \mathrm{s}^{-1}$ \citep{Grinberg_2015}, where $R_{\ast}$ is the radius of the companion star. 
The ratios were computed between the energy bands $0.5$--$1.5/1.5$--$3$ keV (soft colour), and $1.5$--$3/3$--$10$ keV (hard colour), see Fig~\ref{neutral_pl1}.

We then tested a number of physical scenarios in order to find a suitable stellar wind model able to describe the observed colour-colour diagram tracks in Cyg~X-1. 
As our baseline model we chose a primary X-ray continuum modified by a partially covering absorber \citep{Furst_2014,Fornasini_2017,Grinberg_2020}. The adopted model is of the form:

\begin{equation}
\texttt{abs}_{ism} \times \texttt{continuum} \times (f_c \times \texttt{abs}_{wind} + (1-f_c)) \label{partial_cov_model}
\end{equation}
where \texttt{abs$_{ism}$} is the absorption of the interstellar medium (ISM), \texttt{continuum} refers to the unabsorbed primary emission from the X-ray source and \texttt{abs$_{wind}$} is the absorption component associated with the stellar wind. This component blocks a fraction ($f_c$, or "covering fraction") of the X-ray source photons, while only a percentage $1-{f_c}$ reaches the observer modified only by the interstellar medium.

We made use of standard \texttt{XSPEC} models \citep{Arnaud_1996} for the continuum and for the stellar wind to simulate spectra within \texttt{ISIS} \citep{Houck_2000,Houck_2002,Noble_2008}, using the response matrices of observation 201 (see Sect.~\ref{reduction}). We let the parameters of interest of the model vary and, for each combination of parameters, we calculate the corresponding soft and hard colours. In order to speed up the process, for each set of simulations, only one parameter is left free to vary, while the others are kept fixed. This procedure allows us to build simulated colour-colour tracks, which are then compared to the data. We tested different models as described in the following sections. Throughout this work ISM absorption was modelled with \texttt{TBabs}\footnote{\url{https://pulsar.sternwarte.uni-erlangen.de/wilms/research/tbabs/}} using the \texttt{wilm} abundances \citep{Wilms_2000} and the \texttt{vern}  cross-sections\citep{Verner_1996}. The column density of the ISM was kept fixed at the tabulated value of $\sim 0.7 \times 10^{22}\rm{cm}^{-2}$ (\citealt{HI4PIcoll}). Note that in this paper we are interested in testing for variability of the stellar wind component, therefore, in our simulations, we do not investigate intrinsic spectral variations of the X-ray source.


\subsection{Model 1: single power law plus neutral stellar wind}\label{neutral}

We first simulated theoretical colour-colour tracks assuming a power law for the underlying continuum and a partially covering neutral absorber.  
For this simple model, the shape of the colour-colour tracks mainly depends on three parameters: the power law photon index $\Gamma$ of the continuum, the wind covering fraction $f_c$ and the wind column density $N_{\rm{H,w}}$. 

\begin{figure*}[!h]
   \centering \includegraphics[width=1\linewidth]{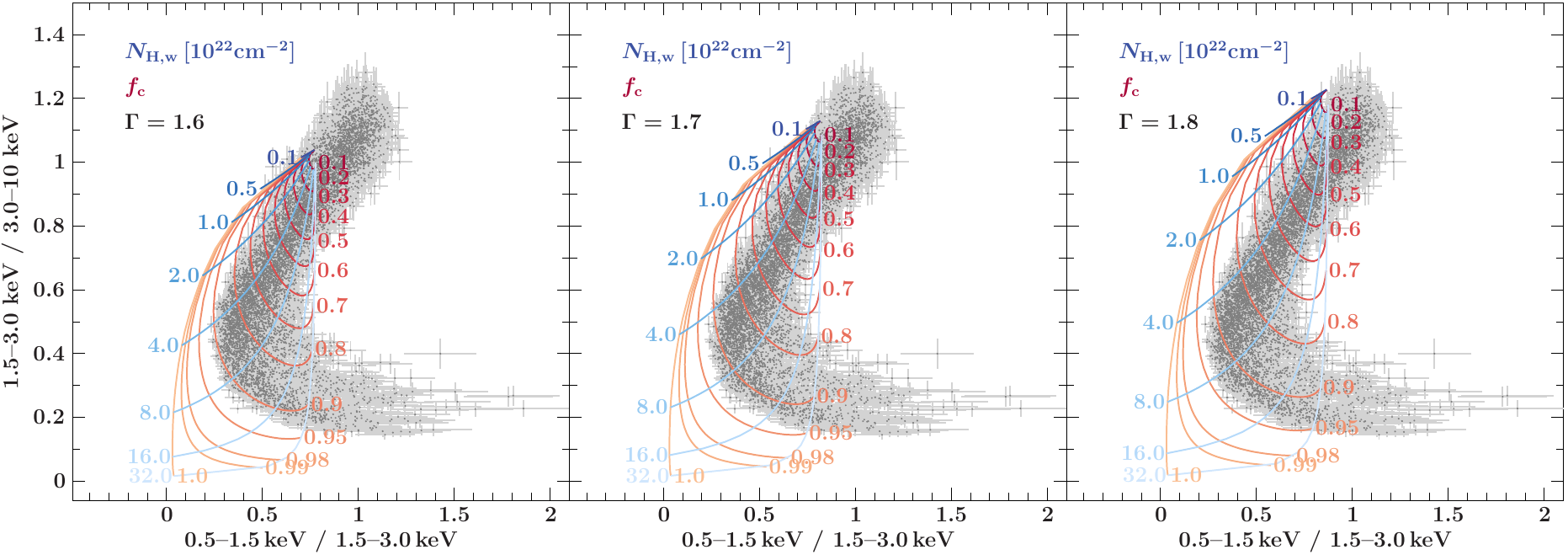}
      \caption{Simulated tracks for a power law (with $\Gamma$ fixed at 1.6, left panel, 1.7, middle panel, and 1.8, right panel) plus a neutral absorber, compared to the observed (gray points) colour-colour diagram tracks of observation 201 of Cyg~X-1. The blue (red) tracks describe changes in the wind $f_c$ ($N_{\rm{ H,w}}$) for a fixed value of $N_{\rm{H,w}}$ ($f_c$).}\label{neutral_pl1}
\end{figure*}

Fig.~\ref{neutral_pl1} shows the resulting simulated tracks overplotted on the colour-colour diagram of observation 201. 
For fixed values of $\Gamma$ and $N_{\rm{H,w}}$, a variable $f_c$ describes the curvature of the blue-cyan tracks shown in Fig.~\ref{neutral_pl1}. Within each track the $f_c$ parameter samples the range of values $f_c=0.1$--$1$ (at steps of 0.1). The different tracks from deep blue to light cyan are obtained by increasing the value of $N_{\rm{H,w}}$ from a minimum value of $N_{\rm{H,w}}=0.1\times10^{22}$ to a maximum value of $N_{\rm{H,w}}=32\times10^{22}\,\rm{cm^{-2}}$ (for higher values of $N_{\rm{H,w}}$ hard colours start to increase again, producing an upward tail which is not observed in the data). On the other hand, for fixed values of $\Gamma$ and $f_c$, a variable $N_{\rm{H,w}}$, within the range $N_{\rm{H,w}}=0.1$--$32\times10^{22}\rm{cm^{-2}}$, describes the curvature of the red-orange tracks shown in Fig.~\ref{neutral_pl1}. The different tracks (from deep red to light orange) are obtained by increasing the value of $f_c$ (from 0.1 to 1). 
We repeated the same simulations for $\Gamma$=1.6, 1.7, and 1.8 (panels from left to right in Fig.~\ref{neutral_pl1}), which match typical values for the hard state \citep[e.g.][]{Joinet_2008, Motta_2009, Gilfanov_2010, Basak_2017, Zhou_2022}. 

We confirm the results of previous studies \citep[e.g.][]{Hirsch_2019,Grinberg_2020} which showed that this simple model fails to describe the observed colour-colour diagram. In particular, a variable $N_{\rm{H,w}}$, while approximating the curvature of the observed track, does not properly reproduce its pointy shape at low hard colours (giving the track a nose-like shape). We observe that a varying $\Gamma$ shifts the tracks vertically. Therefore, in line with previous studies \citep[e.g.][]{Grinberg_2020}, we infer that the data require a more complex modelling of the stellar wind and/or the intrinsic X-ray continuum. Indeed, several X-ray spectroscopic studies highlight the need for a more complex primary continuum for Cyg~X-1 \citep[e.g.][]{Basak_2017,Tomsick_2018} as well as a structured stellar wind made of differently ionised material \citep[][]{Hirsch_2019}.

\subsection{Model 2: structured continuum plus ionised stellar wind}
\label{complexcont}

We then changed our assumptions on the underlying primary continuum and tested a more complex model. 
To this aim we assumed the model presented in \cite{Lai_2022}, which includes a soft and a hard Comptonisation component, as well as direct thermal and reflected emission from the disc. The employed \texttt{XSPEC} model is: \texttt{TBabs} \texttt{$\times$ [diskbb + nthComp + relxillCp]}. In order to properly assess the time-averaged parameters of the continuum we fit a subset of good time intervals (GTIs), selected so as to be the least affected by the intervening stellar wind. To this aim we adopted the same criterion defined in \citealt{Lai_2022}, i.e. we selected data with hard colour $\geq0.95$ and soft colour $\geq0.7$, so as to single out GTIs from the upper, least absorbed region of the colour-colour diagram. Details about the continuum model used in these simulations are reported in Appendix \ref{app_model}.

Fig.~\ref{neutral_multi} shows the colour-colour tracks computed for a neutral gas partially absorbing the adopted structured continuum. As opposed to simulations shown in Fig.~\ref{neutral_pl1}, the new tracks all start right at the center of the observed distribution of data points (upper part of the diagram), meaning that the new continuum model describes the unabsorbed time-averaged spectrum much better than a simple power law. Therefore, hereafter we use this more complex model to describe the continuum in our simulations.

\begin{figure}[!h]
\centering
\includegraphics[width=1\linewidth]{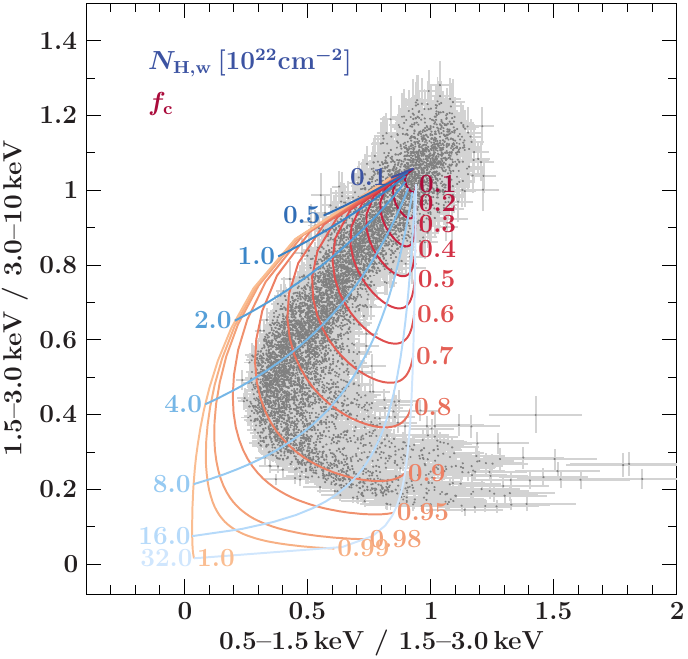}
    \caption{Simulated tracks for the complex continuum absorbed by a neutral wind model, as compared to the colour-colour diagram of Cyg~X-1.}
    \label{neutral_multi}
\end{figure}

\begin{figure*}[!htb]
\centering
\includegraphics[width=1\textwidth]{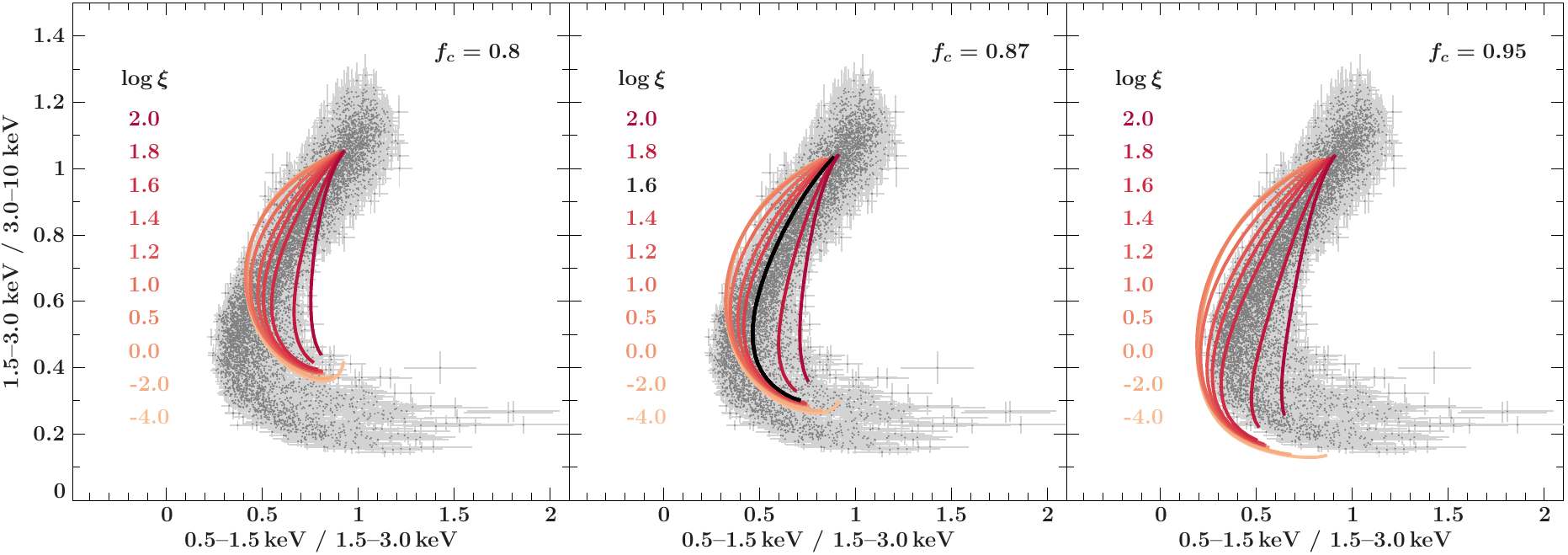}

    \caption{Simulated tracks for a homogeneously ionised absorber (with $\log\xi$ values ranging between -4 and 2) compared to the colour-colour diagram of observation 201. The different panels show the simulated tracks computed for different values of $f_c$: 0.8 (left), 0.87 (middle), 0.95 (right). The black curve in the middle panel represents the one that best describes the data.} 
    \label{warm_multi_3plot}
\end{figure*}

We then considered the effects of an ionised wind on the simulated colour-colour tracks, substituting the \texttt{abs$_{wind}$} component in Eq.~\ref{partial_cov_model} with the 
\texttt{warmabs} (v2.31) analytic photoionisation model \citep{Kallman_2009}.
As in \cite{Grinberg_2020}, we used the standard population files delivered with \texttt{warmabs}, with densities of $10^{12}\,\rm{cm^{-3}}$, typical of stellar wind densities close to the compact object \citep{Lomaeva_2020}.
We first verified how different levels of ionisation modify the simulated tracks. To this aim, we considered the ionisation parameter $\xi=L_X/nr^2$ \citep{Tarter_1969}, where $L_X$ is the ionising luminosity above 13.6 eV, $n$ is the absorbing gas density, and $r$ its distance from the ionising X-ray source. Values of $\log \xi$ between $-4$ and 2 were considered. For each fixed value of $\log \xi$, we let the $N_{\rm{H,w}}$ vary between $0.1$ and $32\times 10^{22}$ cm$^{-2}$, leading to the tracks shown in Fig.~\ref{warm_multi_3plot}. We reproduced the same simulations for three different values of covering factor ($f_c=0.8, 0.87, 0.95$, Fig.~\ref{warm_multi_3plot} from left to right). 
 
By increasing the ionisation parameter the simulated tracks become more pointy, thus more closely resembling the observed tracks. The effect of increasing the covering factor is mostly to increase the range of encompassed hard and soft colours. In particular, for sufficiently high $N_{\rm{H,w}}$, a higher $f_c$ produces more absorption at low-to-intermediate energies, thus causing the absorption dips to be more spectrally hard (lower values of hard and soft colours during the most absorbed stages).
Among the simulated curves, we observe that the one better resembling the observed tracks corresponds to $f_c=0.87$, $\log\xi=1.6$, and $N_{\rm{H,w}}$ varying between $0.1$ and $32 \times 10^{22}$ cm$^{-2}$, as highlighted in black in Fig.~\ref{warm_multi_3plot}, middle panel, thus suggesting mild ionisation for the partially covering wind.

\subsection{Model 3: stellar wind with variable ionisation}\label{stratified}

The simulations shown in Sect.~\ref{complexcont} assumed a constant ionisation parameter throughout the different stages of the absorption dips. Nonetheless, \cite{Hirsch_2019} reported the appearance of less ionised species in the most absorbed stages which corresponds to the lower part of the colour-colour diagram. This suggests the presence of structured clumps, made of differently ionised material. Such results highlight the need to consider changes in the ionisation parameter \citep[][]{Grinberg_2020}.
Therefore, we now account for variations of the stellar wind ionisation parameter as a function of $N_{\rm{H,w}}$. 

\begin{figure*}
\centering
\includegraphics[width=1\linewidth]{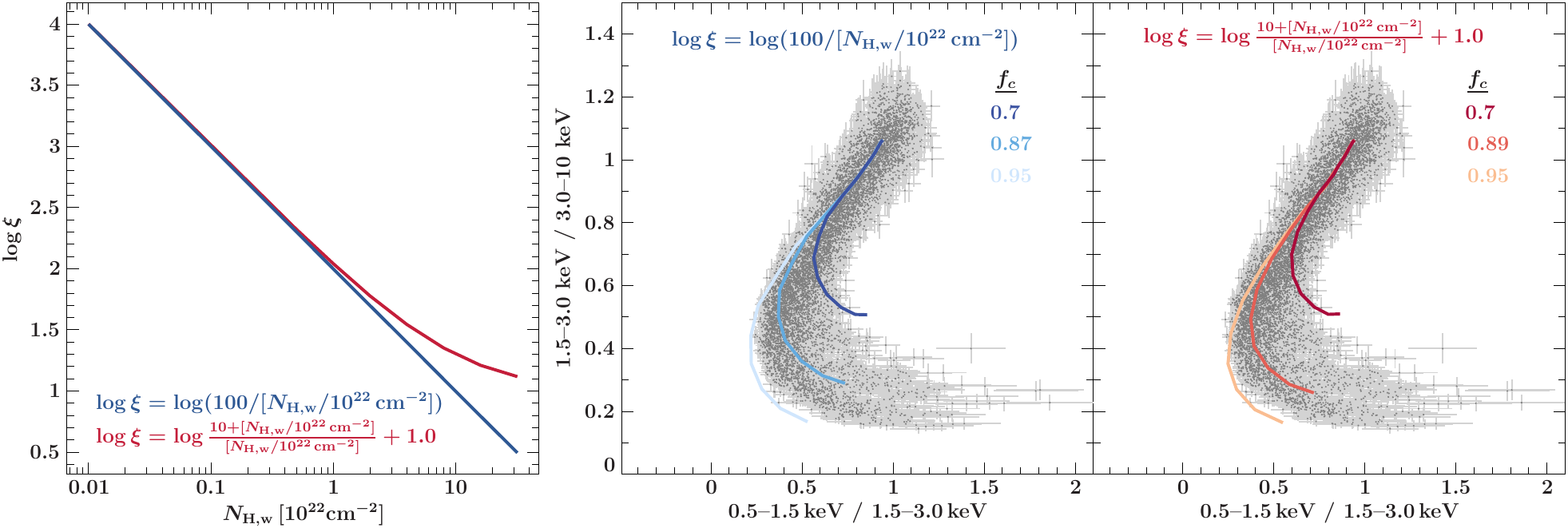}
    \caption{Left panel: the assumed dependencies of the ionisation parameter $\log\xi$ on the column density $N_{\rm{H,w}}$. Middle and right panel: simulated tracks for a warm absorbing gas, assuming $\log\xi=\log(100/[N_{\rm{H,w}}/10^{22}\,\rm{cm}^{-2}])$ (middle panel), and $\log\xi=\log\frac{10+[N_{\rm{H,w}}/10^{22}\,\rm{cm}^{-2}]}{[N_{\rm{H,w}}/10^{22}\,\rm{cm}^{-2}]}+1$ (right panel). The different shades correspond to different values of covering factor as reported in the labels.} 
    \label{warm_merge}
\end{figure*}

Following \cite{Grinberg_2020}, we test two empirical functions to describe the dependence of the ionisation parameter on the column density of the absorbing material. 
The first function is defined as:
\begin{equation}
    \log\xi=\log(A/[N_{\rm{H,w}}/10^{22}\rm{cm}^{-2}]),
\end{equation}
with $A>0$. It is based on the definition of the ionisation parameter and assumes a linear decrease of $\xi$ as a function of $N_{\rm{H,w}}$. 
The second function is defined as:
\begin{equation}
\log\xi=\log\frac{B+[N_{\rm{H,w}}/10^{22}\,\rm{cm}^{-2}]}{[N_{\rm{H,w}}/10^{22}\,\rm{cm}^{-2}]}+C,
\end{equation}
with $B, C>0$. It assumes that the ionisation parameter deviates from a linearly decreasing trend at high column densities, staying relatively high as expected if an additional source of ionisation plays an important role in denser environments \citep[e.g.][]{Feldmeier_1997}. We arbitrarily chose the values of the constant parameters to be $A=100$, $B=10$, and $C=1$. This choice of parameters is such that the two functions match at low column densities $N_{\rm{H,w}}/10^{22}\,\rm{cm}^{-2}<<10$, and start deviating significantly at higher densities, $N_{\rm{H,w}}/10^{22}\,\rm{cm}^{-2}\gtrsim1$.
We let $N_{\rm{H,w}}$ vary between 0.01 and $32\times 10^{22}\,\rm{cm}^{-2}$. Note that sampling down to values of $N_{\rm{H,w}}=0.01\times 10^{22}\,\rm{cm}^{-2}$ allows us to test values of ionisation parameter as high as the highest value allowed by the \texttt{warmabs} model. 
Therefore, the first function spans the range of $\log\xi\sim 0.5$--$4$, while the second function spans the range of $\log\xi\sim1.12$--$4$.
The two empirical functions are shown in Fig.~\ref{warm_merge}.

We built models comprising the complex continuum defined in Sect.~\ref{complexcont} plus each of the two empirical functions describing variable ionisation. 
Since the data in the colour-colour diagrams are not normally distributed, a simple $\chi^2$ minimisation method cannot be used to select the best-fit model. 
In order to find the best-fit model without any prior assumption on the distribution of the data we employed a non-parametric method. 

We used the Kernel Density Estimator (KDE) to infer the Probability Density Function (PDF) of the data in the colour-colour diagram \citep[e.g.][]{Hill1985,Cavecchi_2022}. The KDE is a non-parametric estimator, and its functional form is obtained by combining as many building blocks -- kernels -- as the number of data points. In the KDE a single type of kernel $K$ is used. The choice of the kernel is arbitrary, but for large datasets this choice does not have any significant effect on the final output. Therefore, we used a simple kernel with a Gaussian shape. The kernel $K_i$ centred on the data point $\vec{x}_i$ is defined as:
\begin{equation}
K_i(\vec{x})\propto \exp \left(-\frac{(\vec{x}-\vec{x}_i)^2}{2h^2}\right)
\end{equation}
where $\vec{x}$ is the point in the colour-colour diagram at which the kernel function is estimated. 
The width of the kernel $h$ controls the smoothing. 
The normalised PDF is then computed as:
\begin{equation}
{\rm PDF}(\vec{x})=\frac{1}{n}\sum_{i}^n K_i(\vec{x})
\end{equation}
which sums the contribution of the kernels centred on each data point, $n$ being the total number of data points. 
We implemented the method using the SciPy function  \texttt{gaussian\_kde}, with the optimal kernel width automatically determined by the function (default parameter \texttt{bw\_method='scott'}\footnote{{\url{https://docs.scipy.org/doc/scipy/reference/generated/scipy.stats.gaussian_kde.html}}}).
Using the KDE method we estimated the PDF of the data. Then, for each simulated model we computed the combined likelihood as the product of the values of the probability density at each point of the model. The best-fit model is the one characterised by the highest combined likelihood.

In Fig.~\ref{warm_merge}, we show the best-fit track obtained using the functions $\log \xi=\log(100/[N_{\rm{H,w}}/10^{22}\rm{cm^{-2}}])$ (middle panel) and $\log\xi=\log\frac{10+[N_{\rm{H,w}}/10^{22}\,\rm{cm}^{-2}]}{[N_{\rm{H,w}}/10^{22}\,\rm{cm}^{-2}]}+1$ (right panel) to model the ionised absorber. The best-fit track is obtained in both cases by fitting for the covering factor $f_c$ (between 0.7 and 0.95, at steps of 0.01) parameter while $N_{\rm{H,w}}$ spans the entire range $0.01-32\times 10^{22}\,\rm{cm}^{-2}$ for each fit value of the covering factor. In Fig.~\ref{warm_merge}, we also show the tracks corresponding to $f_c=0.7$ and $f_c=0.95$ and note that outside these limiting values the simulated curves do not intersect the data in the lower part of the diagram. \\
During the least absorbed stages (i.e. at hard colours $\gtrsim$ 0.8), the simulated tracks are not very sensitive to the covering factor, and they all resemble fairly well the observed shape of the track in this part of the diagram. However, at lower values of hard colours, corresponding to the most absorbed phases, the dependence on the covering factor becomes significant. 
The two variable ionisation models considered here describe the data well, with a mild 5\% difference between their inferred probabilities, which does not allow us to prefer either one of the two models. Notably, both functions resemble the pointy shape of the data much better than the homogeneously ionised absorber considered in Sect.~\ref{complexcont}, giving more support to the hypothesis of a structured absorber.
Finally, we observe that the data all lay within the simulated tracks for $f_c=0.7$ and $f_c=0.95$, 
implying that the scatter of data points in this part of the diagram might be in part due to intrinsic variations of the covering factor between ${\sim}0.7-0.95$. We investigate this problem further in the following section.

\section{Time-resolved colour-colour diagrams}\label{time-resolved}

\begin{figure}
\includegraphics[width=1.0\linewidth]{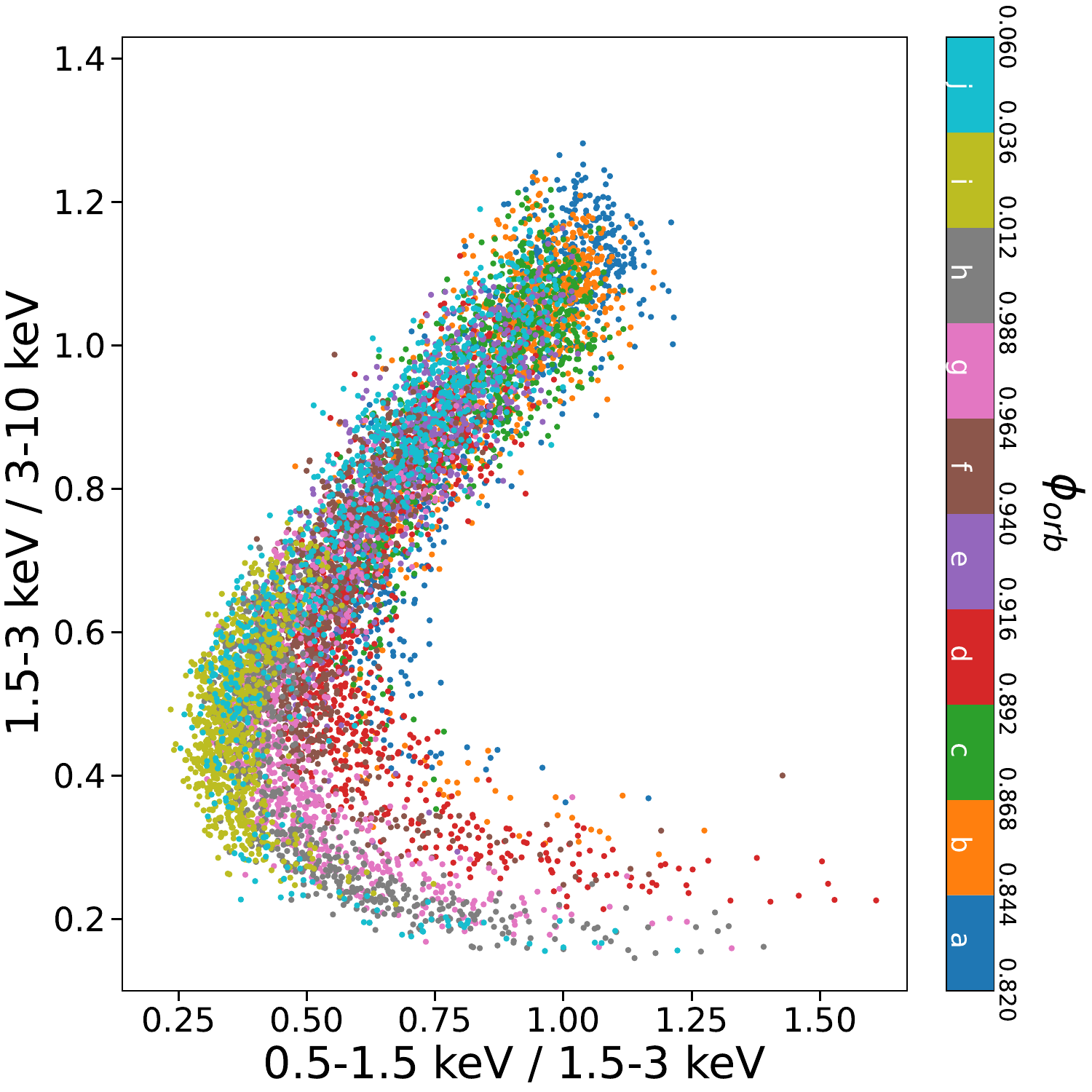}
    \caption{Time-resolved colour-colour diagrams of observation 201, extracted from segments with a duration of 11\,ks (corresponding to ${\sim}0.024$ in orbital phase) and time resolution of 10 s. The resulting ten colour-colour diagrams are overplotted using different colours that correspond to different phase intervals, according to the colour map reported on the right.}
    \label{time_resolved_CCDiagram}
\end{figure}

\begin{figure*}[!h]
\centering 
\includegraphics[width=1\linewidth]{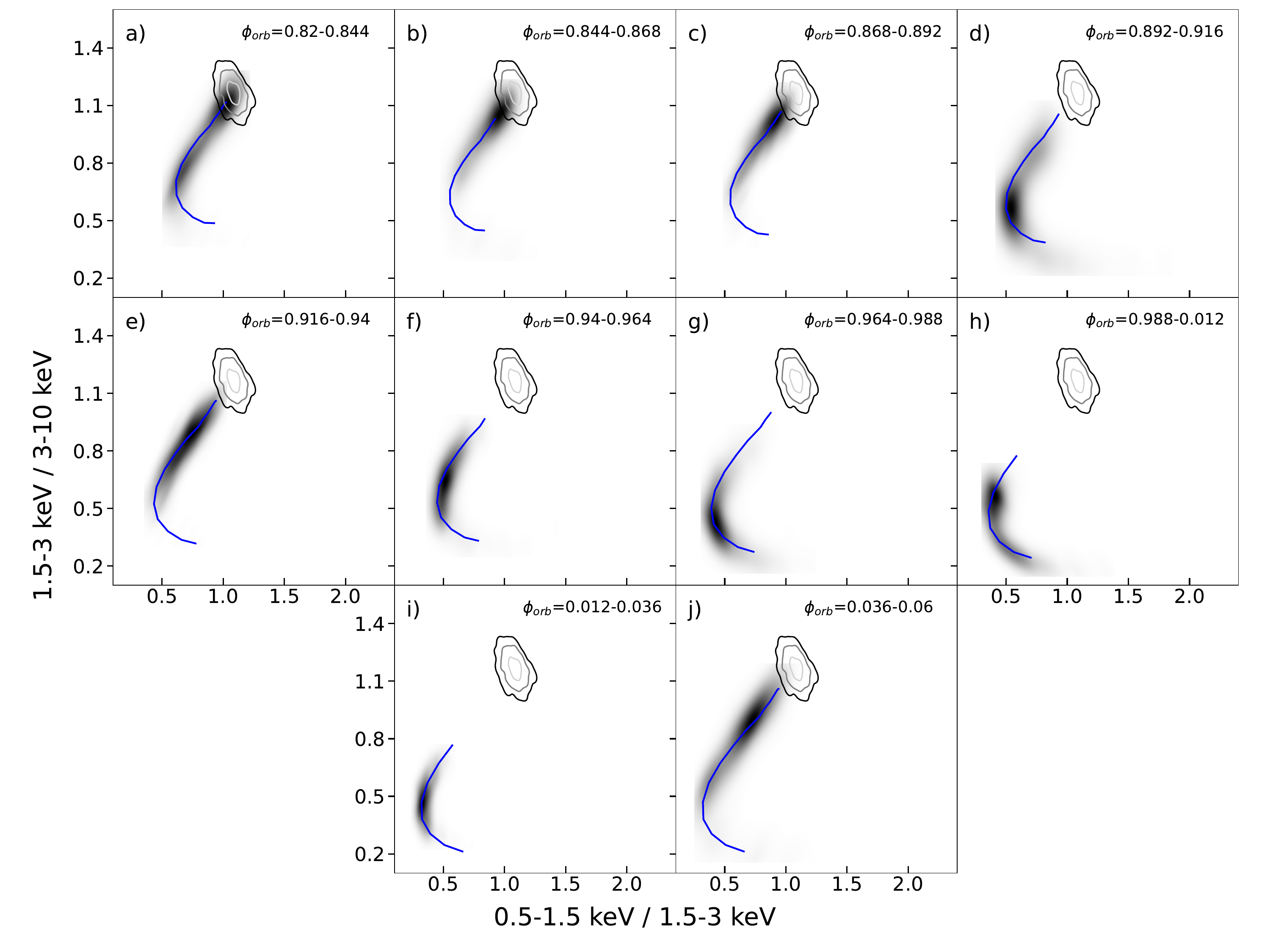}
     \caption{Probability distribution maps of each time-resolved colour-colour diagram obtained using the KDE method. In blue, the best-fit simulated track. The grey-shaded closed curves represent the probability distribution map for orbital phases $\phi_{\mathrm{orb}}=0.43$--$0.46$ at the 99.7\% (in black), 95\% (in grey) and 68\% (in light grey) confidence levels.} 
    \label{PDFmaps}
\end{figure*}

To study the phase-dependent variability of the absorbing column density and covering fraction of the wind, we investigated the temporal evolution of the shape of the colour-colour diagram. 
We divided observation 201 in 10 segments of the same duration of $\sim$$11\,\rm{ks}$ each, corresponding to segments of $\sim$$0.024$ in orbital phase, and produced colour-colour diagrams for each segment. The corresponding tracks are all shown in Fig.~\ref{time_resolved_CCDiagram}, colour-coded based on the range of orbital phases. The plot shows a strong temporal evolution of the tracks. 
The same data are shown in Fig.~\ref{time_resolved_simulated_tracks}, split in separate time-resolved colour-colour diagrams. Using the KDE method (see Sect.~\ref{stratified}), we inferred the PDF of the data in each diagram. The resulting probability distribution maps are shown in Fig.~\ref{PDFmaps}.

The passage at superior conjunction happens at $t\, \sim 86.36\,\rm{ks}$ after the starting time of the observation (see Fig.~\ref{lc_hardness_201}). Therefore, it is contained in panel "h" corresponding to $\phi_{\mathrm{orb}} = 0.988$--$0.012$. 
The fact that the LOS passes through the densest regions of the stellar wind in this phase of the orbit is clearly seen from the shape of the corresponding colour-colour track, whereby \emph{only} the lower region of the diagram is populated (i.e. at hard colours $\lesssim 0.75$).
A similar distribution is observed both at earlier and later phases of the orbit from panel "f" to panel "i" due to strong dipping events occurring several hours before and after superior conjunction (see Fig.~\ref{lc_hardness_201}).

We built simulated tracks for each time-resolved colour-colour diagram in order to constrain the changes in the wind parameters which can cause the observed changes in the colour-colour tracks. 
To this aim, we used the complex continuum model fitted to the spectrum of each time-resolved epoch described in Sect.~\ref{complexcont} and the variable ionisation absorption component described by the empirical function $\log\xi=\log\frac{10+[N_{\rm{H,w}}/10^{22}\,\rm{cm}^{-2}]}{[N_{\rm{H,w}}/10^{22}\,\rm{cm}^{-2}]}+1$, introduced in Sect.~\ref{stratified} (red curve in the left panel of Fig.~\ref{warm_merge}). Although this function cannot be statistically preferred to the other one tested in Sect.~\ref{stratified} (blue curve in the left panel of Fig.~\ref{warm_merge}), our choice is dictated by the willingness to include the effects of additional sources of ionization becoming important in denser environments \citep[e.g.][]{Feldmeier_1997,Sundqvist2018b}, such as the base of the wind, intercepted by the LOS at superior conjunction.
We selected as best-fit tracks those with the highest combined likelihood, as described in Sect.~\ref{stratified}. The best-fit tracks are shown in Fig.~\ref{PDFmaps} (as well as in Fig.~\ref{time_resolved_simulated_tracks}), and the corresponding best-fit covering factors for each time-resolved diagram are listed in Tab.~\ref{tab_wind}. 

We notice that the best-fit tracks do not always extend to high hard colours as much as the data do. This particularly happens in the less absorbed phases of the orbit when the source is out of superior conjunction (i.e. panels a, b, c, and j). This is because the model reaches the maximum allowed value for the ionisation parameter (in our model, the lowest values of $N_{\rm{H,w}}$ correspond to the highest values of $\xi$, Fig.~\ref{warm_merge}, left panel). The upper portion of the diagram not covered by the wind model track is populated by the least absorbed data bins. Therefore, the scattering in the measured hard and soft colours within this region might be completely driven by the intrinsic spectral variability of the source.  
To verify this, we calculated a colour-colour diagram from observation 501, which is one of the least affected by the wind as it was carried out right before inferior conjunction ($\phi_{\mathrm{orb}}=0.5$; see \citealt{Lai_2022}). We note that the CHOCBOX campaign did not catch the passage at inferior conjunction, therefore from this observation we selected the closest orbital phases to it ($\phi_{\mathrm{orb}}=0.43$--$0.46$).
Observation 501 is also consecutive to observation 201. This allows us to rely on the implicit assumption that the intrinsic continuum spectrum of the X-ray source has not changed significantly between the two observations, and that any scatter in the colour-colour diagram would only be due to fluctuations in the parameters of the same continuum model.
Using the KDE method (Sect.~\ref{stratified}), we calculated the probability distribution of soft and hard colours in the diagram. The colour-colour diagram of observation 501 and the corresponding contour plots are reported in Appendix~\ref{app_501} and Fig.~\ref{kde_501_area}. As it can be seen, the data are quite scattered, suggesting intrinsic short term spectral variability of the continuum. In Fig.~\ref{PDFmaps} and \ref{time_resolved_simulated_tracks} we overplot the 99.7\%, 95\% and 68\% confidence contours of observation 501, to mark the area in the diagram where the intrinsic emission from the X-ray source dominates. In other words, in this region the probability for the source to be unabsorbed by the stellar wind is the highest. The data points not reached by the wind model tracks are thus fully consistent with being free from wind absorption.

\begin{figure}
\includegraphics[width=1.0\linewidth]{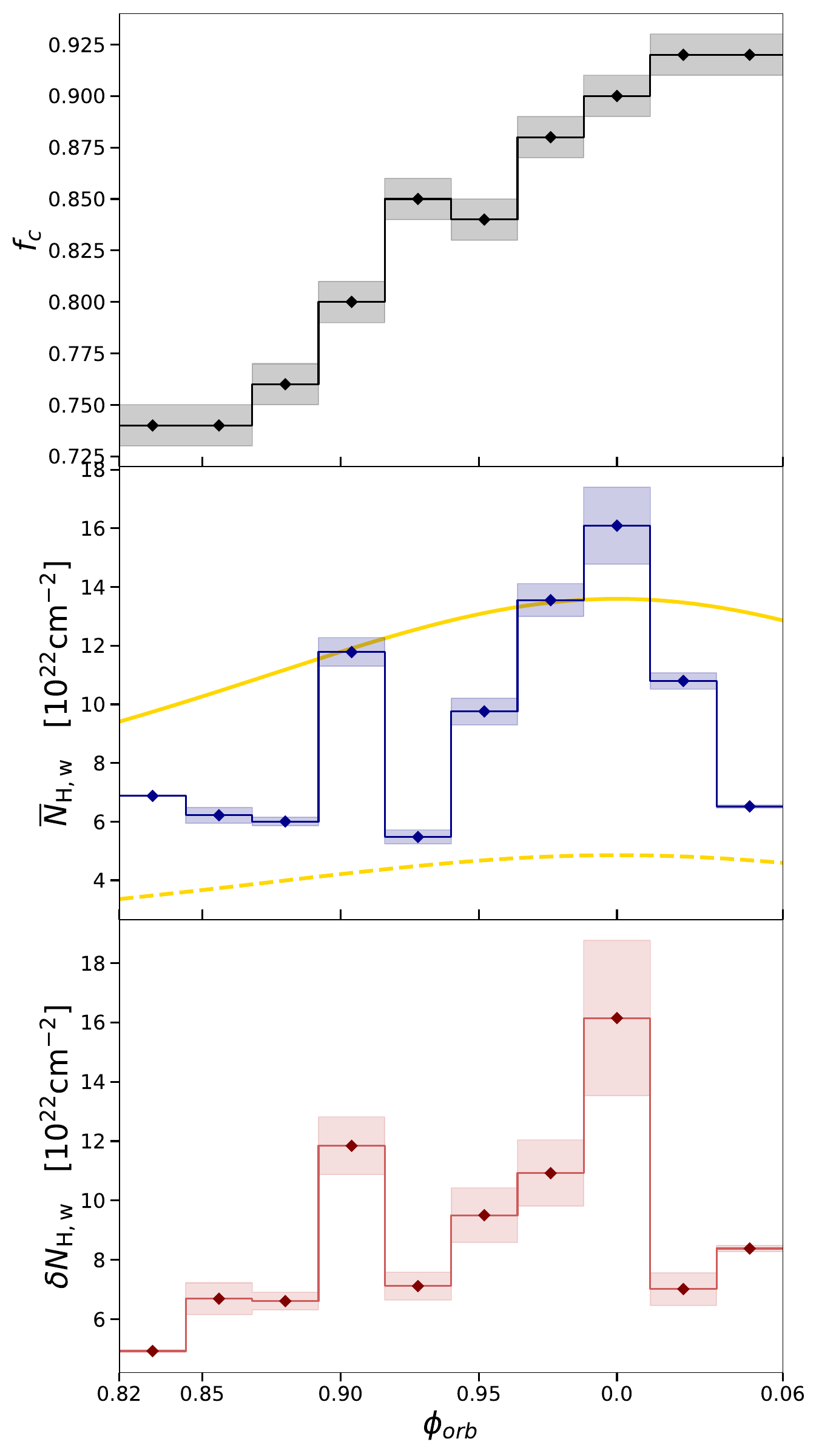}
    \caption{Stellar wind parameters as obtained from the fit of the time-resolved colour-colour diagrams of observation 201, plotted as a function of the orbital phase (see also Tab.~\ref{tab_wind}). The parameters are: the covering factor $f_c$ (upper panel), the mean column density $\overline{N}_{\rm{H,w}}$ (middle panel), and the scatter in column density parameter $\delta N_{\rm{H,w}}$ (lower panel). The shaded areas represent the 68\% confidence contours. The yellow curves are the neutral column density profiles produced by the smooth wind model \citep[][]{ElMellah_2020} at high (solid) and low (dashed) stellar mass loss rates.}
    \label{variability_N}
\end{figure}

From the fit of the time-resolved colour-colour diagrams we infer a steady and significant increase of the covering factor, by a factor $ \sim$$1.2$ between $\phi_{\mathrm{orb}}=0.820$--$0.844$ and $\phi_{\mathrm{orb}}=0.036$--$0.06$ (see Tab.~\ref{tab_wind}). The observed trend is plotted in Fig.~\ref{variability_N} in the upper panel. The maximum value of the covering factor is registered at superior conjunction, but the data suggest this parameter remains high up to at least $\phi_{\mathrm{orb}}=0.06$. 
From the PDFs of Fig.~\ref{PDFmaps} and our best-fit tracks we extracted the average column density $\overline{N}_{\rm{H,w}}$ and its scatter $\delta N_{\rm{H,w}}$, as this parameter is an indicator of the amount of short-term variability of $N_{\rm{H,w}}$ within each epoch. These values are reported in Tab.~\ref{tab_wind} and plotted in Fig.~\ref{variability_N} (middle and lower panels) for each epoch.
Our procedure is this: given the best-fit track, we extract the probability of each point from the KDE, which is thus a function of $N_{\rm{H,w}}$. This probability is a proxy for the number of data points with a given $N_{\rm{H,w}}$. Then, we consider the range of $N_{\rm{H,w}}$ which encloses 68\% of the total (renormalised) probability for the corresponding track, a proxy for the total number of data points. The width of the range gives us $\delta N_{\rm{H,w}}$ and its average value gives us $\overline{N}_{\rm{H,w}}$.
We reckon this approach to be more informative than, for example, extracting the $N_{\rm{H,w}}$ with the highest probability, since the latter method would result in a loss of information, especially for the distributions that show more than one peak. For example, the fit of the colour-colour diagram of panel "a" of Fig.~\ref{PDFmaps} would return a low $N_{\rm{H,w}}$, thus missing the information associated with the second peak of the PDF.
Our procedure provides also a straightforward way to obtain estimates of the confidence intervals for each pair of $\overline{N}_{\rm{H,w}}$ and $\delta N_{\rm{H,w}}$ values. We now consider the two tracks corresponding to the extrema of the 68\% confidence level of $f_c$ and repeat the procedure obtaining new confidence level of $N_{\rm{H,w}}$. To be conservative, we take the overall maximal and minimal confidence level and from these obtain the intervals of $\overline{N}_{\rm{H,w}}$ and $\delta N_{\rm{H,w}}$.

We observe significant variations in both the $\overline{N}_{\rm{H,w}}$ and the $\delta N_{\rm{H,w}}$ parameters. While the two show the same trend of variations (when the average column density increases, its scatter also increases), their trend differs from that of $f_c$. In particular, both the $\overline{N}_{\rm{H,w}}$ and the $\delta N_{\rm{H,w}}$ show two peaks, the second one happening at superior conjunction, when also $f_c$ reaches its maximum value. However, when $f_c$ is still at its maximum, $\overline{N}_{\rm{H,w}}$ and $\delta N_{\rm{H,w}}$ drop by a factor of $\sim$$2.5$ and $\sim$$2$, respectively. This is readily visible in panel "j" of Fig.~\ref{PDFmaps} and Fig.~\ref{time_resolved_simulated_tracks}, where the data populate again the region of high hard colours of the colour-colour diagram (as in panels "a--e") while also stretching down to the lowest registered hard colours. 
Finally, it is worth noticing that the measured values of $\overline{N}_{\rm{H,w}}$ (between $\sim$$6\times 10^{22}\,\rm{cm}^{-2}$ and $\sim$$16\times 10^{22}\,\rm{cm}^{-2}$) all fall in the regime where our model predicts additional ionisation in high density environments to become significant (see Fig.~\ref{warm_merge}, left panel). In this regime, strong variations of $\overline{N}_{\rm{H,w}}$ along the dips are not mirrored by equally strong variations of the ionisation parameter, which remains constrained between $\log\xi\sim$$1.2$--$1.4$.

\begin{table}
\setlength{\tabcolsep}{5pt}
\renewcommand{\arraystretch}{1.5}
\caption{Parameters obtained from the fits of the time-resolved colour-colour diagrams of observation 201. The table reports the ranges of orbital phases corresponding to each time-resolved diagram. The average column density $\overline{N}_{\rm{H,w}}$ and its scatter $\delta N_{\rm{H,w}}$ are in units of $10^{22}\ {\rm cm^{-2}}$.}
\label{tab_wind}      
\centering          
\begin{tabular}{c c c c}
\hline
$\phi_{\mathrm{orb}}$ & $f_c$ & $\overline{N}_{\rm{H,w}}$ & $\delta N_{\rm{H,w}}$\\ 
\hline                    
   $0.820-0.844$ & $0.74\pm0.01$   & $6.88\pm{0.02}$   & $4.93\pm{0.03}$\\
   $0.844-0.868$ & $0.74\pm0.01$   & $6.22\pm{0.27}$   & $6.70\pm{0.53}$\\
   $0.868-0.892$ & $0.76\pm0.01$   & $6.00\pm{0.15}$   & $6.6\pm{0.3}$\\
   $0.892-0.916$ & $0.80\pm0.01$   & $11.78\pm{0.48}$  & $11.84\pm{0.97}$\\
   $0.916-0.940$ & $0.85\pm0.01$   & $5.48\pm{0.23}$   & $7.12\pm{0.47}$\\
   $0.940-0.964$ & $0.84\pm0.01$   & $9.76\pm{0.46}$   & $9.51\pm{0.91}$\\
   $0.964-0.988$ & $0.88\pm0.01$   & $13.55\pm{0.56}$  & $10.92\pm{1.11}$\\
   $0.988-0.012$ & $0.90\pm0.01$   & $16.09\pm{1.31}$  & $16.15\pm{2.62}$\\
   $0.012-0.036$ & $0.92\pm0.01$   & $10.80\pm{0.28}$  & $7.02\pm{0.55}$\\ 
   $0.036-0.060$ & $0.92\pm0.01$   & $6.52\pm{0.06}$   & $8.38\pm{0.11}$\\ 

\hline           
   
\end{tabular}
\end{table}

\subsection{On the soft tail of colour-colour tracks}\label{tail}

Before discussing our results, we note that none of the models used to fit the colour-colour diagrams (Sect.~\ref{Colour-colour diagrams} and Sect.~\ref{time-resolved}) produces tracks that can explain the extended tail at high values of soft colours characterising the most absorbed stages of the dips (lower part of the colour-colour diagram, see e.g. Fig.~\ref{warm_merge} and Fig.~\ref{time_resolved_simulated_tracks}).

Such high soft colours indicate the presence of an additional soft component which becomes significant in the deeper parts of the dips. The most likely candidates for explaining this behaviour are the contribution from emission lines produced by the wind material, and/or from a dust scattering halo \citep[e.g.][]{Nowak_2011}.
The former possibility is supported by results presented in \cite{Hirsch_2019}, who revealed the presence of an emission line spectrum from the photoionised plasma around the BH, emerging only when the X-ray source is highly absorbed (down in the dips). Additional contribution from a dust scattering halo cannot be excluded. Indeed, source photons intercepted by foreground interstellar dust on the LOS are scattered away, thus dimming the source, but interstellar dust grains outside the LOS can redirect photons back to the observer, producing a halo around the source. Given the energy dependence of the dust scattering cross-section \citep{Corrales_2016} soft X-ray photons will have more chance to be scattered back, thus producing an excess of soft X-ray flux in the halo \citep{Maeda_1996}.

In order to check for the presence of a dust scattering halo component in the {\it XMM-Newton} EPIC-pn data of Cyg~X-1, we verified whether the spectrum significantly softens when extracting photons from increasingly more external regions, as described in \cite{Jin_2017}. It is worth noticing that the timing mode does not allow us to select regions totally unaffected by a dust scattering halo since all pixels in the same column lose their spatial information along RAWY. Nonetheless, if a scattering halo is present and the X-ray source is bright, the inner columns will still be harder than the external columns due to the X-ray source dominating the central columns and the halo dominating the external ones. 

We extracted spectra in the $2$--$10$ keV band during the time interval $\sim$$82$--$94\,\rm{ks}$ which comprises the passage at superior conjunction. Indeed the emission from the halo is expected to be more significant when the X-ray source is highly absorbed \citep[e.g.][]{Jin_2018}, thus in the deepest stages of the dips.
We selected three detector regions at increasingly higher distance from the central pixels (and discarding the central RAWX$=[36:39]$ to avoid pile-up effects, Sect.~\ref{reduction}): RAWX $= [32:35]$--$[40:43]$ covering an angular diameter D $=8"$--$16"$ from the central pixel, RAWX $= [30:32]$--$[43:46]$ (D $=16"$--$32"$) and RAWX $= [27:30]$--$[46:49]$ (D $=32"$--$44"$). 
A gradual softening of the $\Gamma$ parameter is observed, up to $\sim$$10\%$ in the most external region. In particular, $\Gamma=1.42\pm0.01$ for RAWX $=[32:35]$--$[40:43]$, $\Gamma=1.54\pm0.01$ for RAWX $=[30:32]$--$[43:46]$, and $\Gamma=1.58\pm0.01$ for RAWX $=[27:30]$--$[46:49]$. 
This gradual softening hints at the presence of a dust scattering halo component. We conclude that the dust scattering halo can potentially contribute to the soft tail observed in the colour-colour tracks. 
Nonetheless, we notice that both a dust scattering halo and the emission line component from the wind would contribute the most when the source is highly absorbed. Thus we expect a soft tail to be particularly prominent when the uppermost part of the colour-colour diagram is not populated. However, all the observed time-resolved colour-colour diagrams follow these expectations (more visible in Fig.~\ref{time_resolved_simulated_tracks}), except the one at $\phi_{\mathrm{orb}}=0.012$--$0.036$ (see panel "i" of Fig.~\ref{time_resolved_simulated_tracks}) which does not show a prominent soft tail despite being highly absorbed ($\overline{N}_{\rm{H,w}}\sim 10.80 \times 10^{22}\,\rm{cm}^{-2}$ and $f_c\sim0.92$). We suggest this might be due to a delayed response of the illuminated dust halo or emission line region to a change of the irradiating flux from the X-ray source (see Fig.~\ref{lc_hardness_201}).

\begin{figure}[!h]
\centering 
\includegraphics[width=1\linewidth]{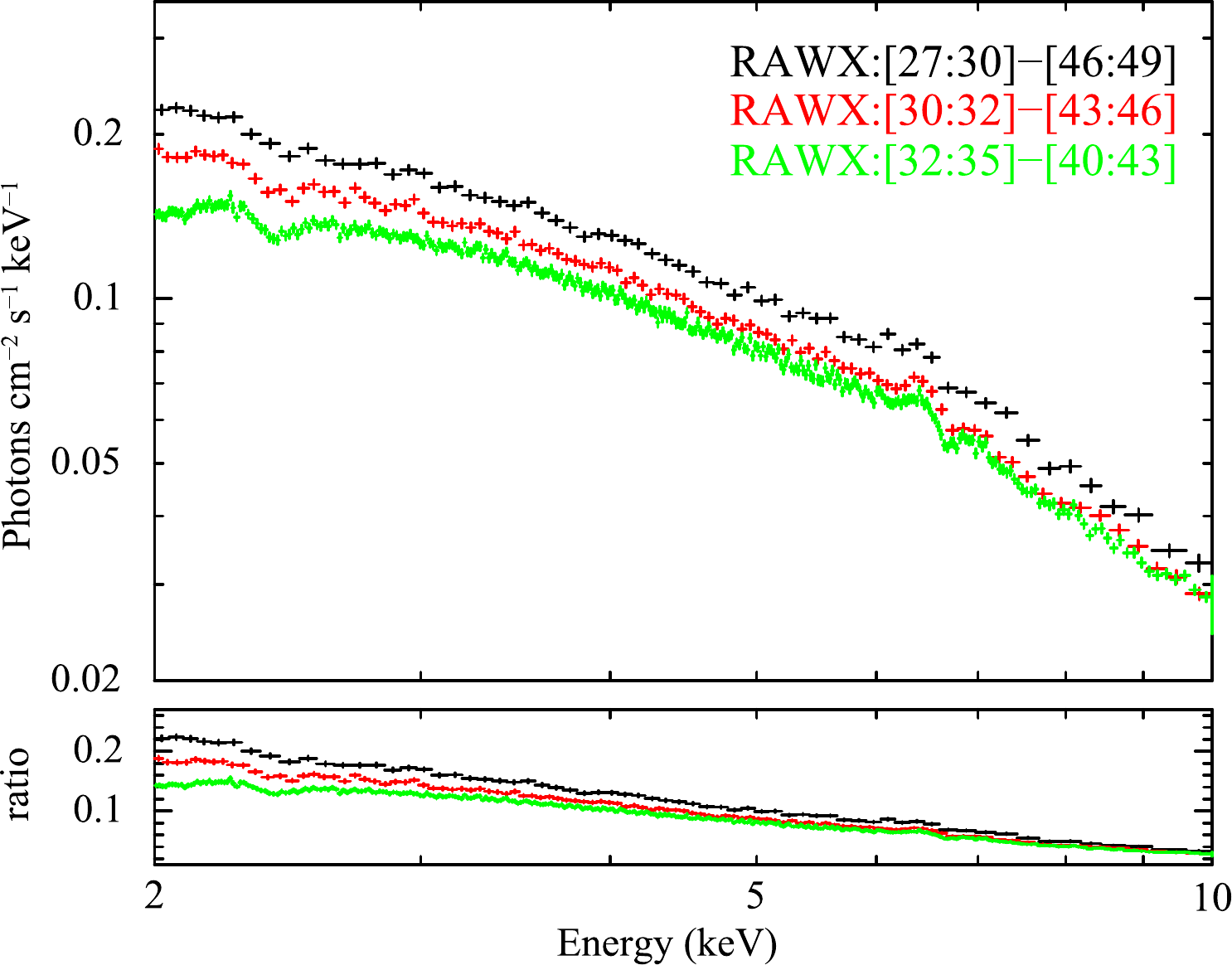}
     \caption{Unfolded (against a constant model fixed at 1) 
     $2$--$10$ keV EPIC-pn spectra (upper panel) extracted in the time interval $t\sim82$--$94$ ks, catching the passage at superior conjunction. The chosen RAWX selections for each spectrum are reported in the labels. The softening of the spectra suggests that there is contribution from a dust scattering halo, easily noticeable in the ratios of the unfolded spectra (bottom panel).
} 
    \label{dust_scattering_test}
\end{figure}


\section{Discussion}\label{discussion}

Several studies demonstrated the complexity of the stellar wind in Cyg~X-1 \citep[e.g.][]{Grinberg_2015,Miskovicova_2016,Hirsch_2019,Grinberg_2020}. 
The wind appears to be formed of over dense clumps (e.g. \citealt{Owocki_1988,Feldmeier_1997,Oskinova_2012,Sundqvist_Owocki_2013}) crossing our LOS to the X-ray source, producing stochastic variability which typically manifests itself as absorption dips in the X-ray light curves, and adds up to the intrinsic variability produced in the inner accretion flow \citep[][]{Lai_2022}. Absorption dips occurrence and concurrent absorption variability are more prominent near superior conjunction \citep[e.g.][]{Balucinska_2000,Lai_2022}, when the companion star is between the observer and the BH ($\phi_{\mathrm{orb}}=0$), as in these phases the LOS crosses deeper wind layers. 
 
In this work, we presented a characterisation of the stellar wind properties in Cyg~X-1, through the analysis and modelling of the colour-colour diagrams of the source during a passage at superior conjunction (between orbital phases 0.82 and 0.06, see Fig.~\ref{fig:chocbox} and Fig.~\ref{lc_hardness_201}), with the X-ray source in its hard state. The prominence of absorption dips in this specific state, rather than in softer states, are explained as due to the X-ray source not being strong enough to fully ionise the wind \citep[e.g.][]{Miskovicova_2016,Grinberg_2015}. 
In addition to previous studies, we employed a non-parametric fitting method based on the kernel density estimation analysis \citep{Hill1985}, which allowed us to select the model that best describes the observed colour-colour diagrams and constrain the physical parameters of the stellar wind (Sect.~\ref{stratified} and Sect.~\ref{time-resolved}).

\subsection{Modelling of colour-colour diagrams and evolution of stellar wind parameters}
We confirmed (see also \citealt{Grinberg_2020} who analysed a Chandra 2004 observation of the source covering similar orbital phases) that the characteristic "pointy" or "nose-like" shape of the colour-colour tracks requires the absorbing gas to be partially ionised (Sect.~\ref{complexcont}), in agreement with independent results from high resolution spectroscopy \citep[e.g.][]{Hanke_2009,Miskovicova_2016}. 
In particular, \cite{Hirsch_2019} revealed different levels of ionisation at different absorption stages. This suggests an  (unknown) functional dependence of $\log\xi$ on $N_{\rm{H,w}}$ \citep{Grinberg_2020}. We tested two empirical functions (Sect.~\ref{stratified}), one assuming a simple linear scaling and the other allowing for additional sources of ionisation at high column densities (Fig.~\ref{warm_merge}). Although the data do not allow us to statistically prefer either of the two models, we notice that simulations show that collisions in the ambient stellar wind occurring in denser regions (high $N_{\rm{H,w}}$ regime) are expected to produce significant X-ray emission \citep[][]{Feldmeier_1997,Sundqvist2018b}. This emission can contribute to further ionise the gas.

\begin{figure}
\includegraphics[width=1.0\linewidth]{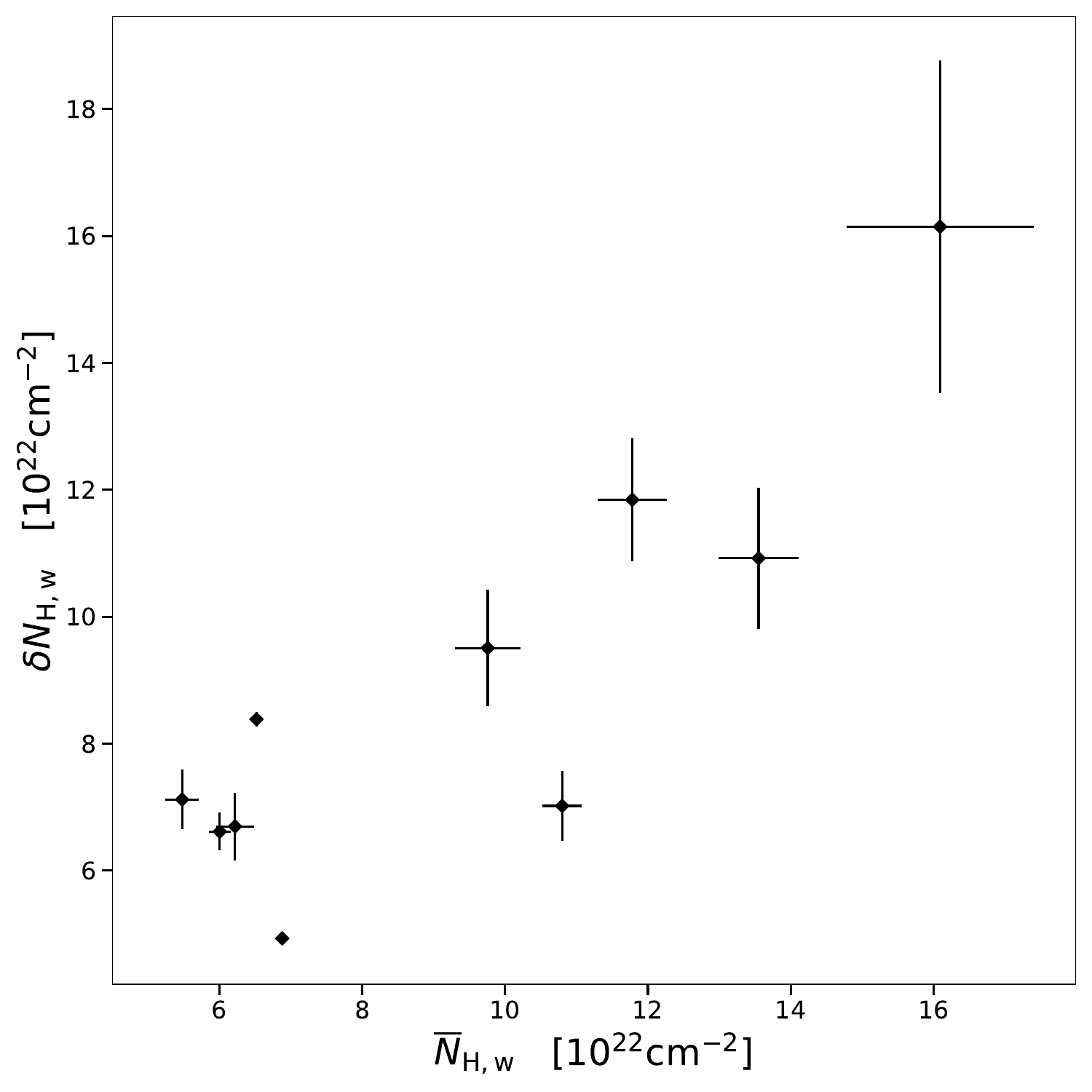}
    \caption{The relation between the measured $\overline{N}_{\rm{H,w}}$ and $\delta N_{\rm{H,w}}$ parameters, respectively sampling long-term and rapid variations of the column density parameter of the stellar wind (the values are listed in Tab.~\ref{tab_wind}).}
    \label{NH_deltaNH}
\end{figure}

In order to characterise the evolution of the wind as a function of the orbital phase around superior conjunction, we extracted time-resolved colour-colour diagrams, and fit them with a continuum plus variable ionisation model (see Sect.~\ref{time-resolved}).
The fits show a steady increase of the covering factor $f_c$ by a factor $\sim$$1.2$ between $\phi_{\mathrm{orb}}\sim0.8$ and $\phi_{\mathrm{orb}} \sim 0$, which remains high also after superior conjunction (up to at least $\phi_{\mathrm{orb}}\sim0.06$, Fig.~\ref{variability_N} top panel and Tab.~\ref{tab_wind}). On the other hand, the mean column density $\overline{N}_{\rm{H,w}}$ shows two peaks, one at $\phi_{\mathrm{orb}}\sim0.9$ and the other at superior conjunction. After superior conjunction, while $f_c$ is still high, $\overline{N}_{\rm{H,w}}$ drops to a minimum (Fig.~\ref{variability_N} middle and Tab.~\ref{tab_wind}). We also measured the scatter parameter $\delta N_{\rm{H,w}}$, which quantifies the amount of variability of $N_{\rm{H,w}}$ within each time-resolved colour-colour diagram (Fig.~\ref{variability_N} bottom panel and Tab.~\ref{tab_wind}). This parameter follows the same trend displayed by $\overline{N}_{\rm{H,w}}$ (see also Fig.~\ref{NH_deltaNH}). 
Given the sampling adopted for the extraction of time-resolved colour-colour diagrams (10 segments of $\sim$11 ks), while the inferred $\overline{N}_{\rm{H,w}}$ parameter probes relatively long term modulations, the $\delta N_{\rm{H,w}}$ parameter probes more rapid variability (between 10 s, the time bins used to build the colour-colour diagrams, and $\sim$$11$ ks) likely driven by the smallest-scale inhomogeneities in the wind. 
Our study hints at a one-to-one relation between the amount of absorption at a given phase and its associated rapid variability, i.e. the higher the $\overline{N}_{\rm{H,w}}$, the higher its scatter $\delta N_{\rm{H,w}}$ (see Fig.~\ref{NH_deltaNH}). This is reminiscent of the ``rms-flux'' relation characterising the stochastic flux variability in compact object systems \citep[e.g.][]{Uttley2005,Scaringi2012}, which implies a log-normal distribution of the flux and requires variability on different time scales to combine multiplicatively \citep[][]{Uttley2005}. Since the size of the clumps influences the observed column density, confirmation of such correlation might have implications on the distribution of clump sizes and on the way they combine to form bigger clumps.

\subsection{Stellar wind mass loss rate and clump mass estimations}

The variations of the absorbing column density stem from two aspects. First, we expect a periodic variation due to the orbital motion of the BH around its stellar companion. This smooth component can be described by the absorption from a smooth wind, but it falls short at capturing the stochastic variability on timescales much shorter than the orbital period. The latter comes from clumps intercepting the LOS between the observer and the BH \citep[][]{ElMellah_2020}. The average column density time series match the smooth wind profile of same mass loss rate, while the dispersion above and below tell us about the clumps' properties. 

The data considered in this paper cover only a small portion of the orbital period, so they cannot be used to constrain the periodic variations of column density. Instead, we plotted in the middle panel in Fig.~\ref{variability_N}, for two different stellar mass loss rates, the periodic component of the column density due to the orbital motion (yellow lines). It was computed based on the smooth and spherically-symmetric stellar wind model of \citeauthor{ElMellah_2020} (\citeyear{ElMellah_2020}): it accounts for all the material along the LOS to the BH, regardless of its ionization state. Therefore, it is an estimate of the neutral column density, necessarily higher than the measurable ionised column density. If we neglect the orbital eccentricity and assume a $\beta$-law for the wind velocity profile, the shape of the neutral column density curve only depends on the orbital inclination $i$, on the ratio of the orbital separation $d$ to the stellar radius $R_{\ast}$ and on the $\beta$ exponent of the velocity profile. We take $i=27^\circ$ and $d/R_{\ast}=2.3$, which correspond to the values given by \cite{MillerJones_2021}, and $\beta=1.5$, a representative value for UV-line driven winds of O supergiants \citep{RubioDiez_2022}. Owing to the low inclination of Cyg~X-1, the peak-to-peak variation remains modest across the orbit and no significant changes are expected for realistic values of $d/R_{\ast}$ and $\beta$. On the other hand, the scale of the neutral column density curve is set by $\dot{M}_{\ast}/(m_p v_{\infty} R_{\ast})$, where $\dot{M}_{\ast}$ is the stellar mass loss rate and $m_p$ is the proton mass. We use the values reported by \cite{MillerJones_2021} to set $v_{\infty}=2100\ {\rm km/s}$ and $R_{\ast}=22\ R_{\odot}$. Through $H\alpha$ diagnostics, \cite{Gies_2003} reported a stellar mass loss rate $\dot{M}_{\ast}\sim2.6\times10^{-6}\ {\rm M_{\odot} \, yr^{-1}}$ in the hard state. However, they do not take ionization into account, and they rely on the stellar parameters used by \cite{Herrero_1995}, which do not correspond to the updated values. Therefore, there is a risk that this mass-loss rate is underestimated. And indeed, when plotted over our data points, the neutral column density profile with $\dot{M}_{\ast}\sim 2.6\times10^{-6}\ {\rm M_{\odot} \, yr^{-1}}$ (dashed yellow line) lies below the data points. However, since 
the profiles, in Fig.~\ref{variability_N}, produced by the smooth wind model of \citeauthor{ElMellah_2020} (\citeyear{ElMellah_2020}) represent the neutral column density while the data points stand for the absorbing column density (i.e. only the material at ionization levels low enough to contribute to absorption), the smooth profile should be an upper limit. Therefore, we think a higher stellar mass loss rate, for instance $\dot{M}_{\ast}\sim7\times10^{-6}\, {\rm M_{\odot} \, yr^{-1}}$ (solid yellow line), is more compatible with the column density we measure at superior conjunction. Measurements of $\overline{N}_{\rm{H,w}}$ over multiple consecutive orbits are needed to obtain better constraints. 

Moreover, both the $\overline{N}_{\rm{H,w}}$ and the $\delta N_{\rm{H,w}}$ parameters show significant short-term variability (Fig.~\ref{variability_N}). \cite{ElMellah_2020} presented a model that explores the impact of the wind clumpiness on the rapid time-variability of its column density, and describes how this observed variability can be linked to the clumps' physical properties, such as the size and the mass. 
In the clumpy wind model of \cite{ElMellah_2020}, the clumps' properties are directly linked to the scatter parameter $\delta N_{\rm{H,w}}$.
This parameter is expected to be maximum around superior conjunction, as indeed observed (Tab.~\ref{tab_wind} and Fig.~\ref{variability_N}, lower panel). The model also predicts the scatter to be an excellent tracer of the ratio $\sqrt{m_{cl}}/R_{cl}$, where $m_{cl}$ and $R_{cl}$ are the mass and radius of the clumps. In \cite{Lai_2022} we measured the minimum timescales at which the passage of clumps causes significant excess X-ray variability in the power spectra of the source at superior conjunction. These timescales correspond to a clump radial size of $\gtrsim10^{-4}\ R_{\ast}$. 
Using this as the smallest clumps radius, the measured maximum value of $\delta N_{\rm{H,w}}$ (Tab.~\ref{tab_wind}), and the above estimate for the mass loss rate, and plugging them into equation 18 of \cite{ElMellah_2020} we infer $m_{cl}\sim10^{17}\ {\rm g}$ as an estimate for the characteristic mass of the clumps (for $v_{\infty}=2100 \ {\rm km/s}$ and $R_{\ast}=22\ R_{\odot}$). This estimate is in agreement with values inferred from radiative hydrodynamical simulations \citep[][]{Sundqvist_2018}, as well as from spectral analysis \citep[][]{Haerer_2023}.

\subsection{The soft colour tail}
The observed colour-colour tracks also reveal the presence of a soft emission component in the most absorbed stages, producing an extended soft-colour tail. We verified (Sect.~\ref{tail}) that the analysed {\it XMM-Newton} data are consistent with the presence of significant contribution from a dust scattering halo \citep[e.g.][]{Jin_2017}. However, additional contribution from the wind itself in the form of emission lines from the diffused photoionised gas around the BH is also expected \citep[][]{Hirsch_2019}. Indeed, both contributions from a dust scattering halo and the photoionised emitting plasma are expected to be prominent when the primary radiation from the X-ray source is significantly blocked by the absorbing material, i.e. in the deepest stages of the dips. A more complex modelling (which is beyond the scope of this paper) would allow the two contributions to be disentangled, and might in turn be useful to put stronger constraints on wind models, as well as probing the interstellar dust towards Cyg~X-1.

As a final remark, it is worth noting that additional spectral changes extrinsic to the wind, i.e. intrinsic changes of the hard X-ray source properties, have been neglected in our fits of time-resolved colour-colour diagrams. For example, such changes might be due to time variations of the spectral index $\Gamma$, as induced by variations of either the temperature or the optical depth of the Comptonising gas (or both).  
In Sect.~\ref{time-resolved}, we verified that the scattering in hard and soft colours in the uppermost, unabsorbed part of the diagrams are most likely driven by spectral variability. \cite{Mastroserio_2021} showed that $\Gamma$ variability of a few percents is plausible in hard state sources with $\sim$$10\%$ fractional rms variability, typical of the hard state. \cite{Skipper_2013} found a larger $\Gamma$ variability ($\sim$$10\%$) for the same source and spectral state but how these spectral changes influence the most absorbed parts of the diagram is yet to be tested. 


\section{Conclusions}\label{conclusion}

In this paper we presented an analysis of the colour-colour diagrams of Cyg~X-1 during a passage at superior conjunction, with the aim of constraining the properties of the stellar wind. 
We employed the KDE method to select wind models that best fit the colour-colour diagrams. This allowed us to overcome the problem of dealing with data which are not normally distributed. 
The main results are the following:

\begin{itemize}

\item We found that the model that best describes the characteristic "pointy" or "nose-like" shape of the diagrams implies the wind to be partially ionised.

\item We revealed a strong temporal evolution of the colour-colour diagrams around superior conjunction. Our fits suggest this evolution to be strongly influenced by concurrent variations of the column density and covering factor of the wind.

\item Both the column density and the covering factor are maximum at superior conjunction (since the LOS crosses deeper wind layers), but their overall variations follow different trends. 

\item We report a one-to-one scaling between long-term ($>$ 11 ks) and rapid (between 10 s and 11 ks) variations of the column density (reminiscent of the ``rms-flux'' relation characterizing stochastic variability in compact objects). The existence of such correlation might have implications on the distribution of the wind clumps and on the way they combine into bigger clumps.

\item Using the clumpy wind model proposed in \cite{ElMellah_2020} we estimated a wind mass loss rate of $\dot{M}_{\ast}\sim7\times10^{-6}\ {\rm M_{\odot} \, yr^{-1}}$ and a characteristic clump mass of $m_{cl}\sim10^{17}\ {\rm g}$.
\end{itemize}

Future applications of this analysis approach will require the sampling of longer stretches of the binary orbit and the application of more complex models (e.g. including contribution from scattered wind emission and from the dust scattering halo).

\begin{acknowledgements}
      This work is based on observations obtained with {\it XMM–Newton}, an ESA science mission instrument, and contributions directly funded by ESA Member States and NASA.  
      EVL thanks Alex Markowitz, Piotr Życki and Maura Pilia for useful discussions.
      The research leading to these results has received funding from the European Union's Horizon 2020 Programme under the AHEAD2020 project (grant agreement n. 871158). EVL and AR were partially supported by the Polish National Science Centre, grant no. 2021/41/B/ST9/04110.
      EVL and MB are supported by the Italian Research Center on High Performance Computing Big Data and Quantum Computing (ICSC), project funded by European Union - NextGenerationEU - and National Recovery and Resilience Plan (NRRP) - Mission 4 Component 2 within the activities of Spoke 3 (Astrophysics and Cosmos Observations). 
      BDM acknowledges support via Ram\'on y Cajal Fellowship (RYC2018-025950-I) and the Spanish MINECO grant PID2022-136828NB-C44. 
      YC acknowledges support from the grant RYC2021-032718-I, financed by MCIN/AEI/10.13039/501100011033 and the European Union NextGenerationEU/PRTR. BDM, YC, GS, and JJ thank the Spanish MINECO grant PID2020-117252GB-I00 and the AGAUR/Generalitat de Catalunya grant SGR-386/2021. 
      MC acknowledges support from the ``Universitas Copernicana Thoruniensis In Futuro'' project nr POWR.03.05.00-00-Z302/17 and  from the MNiSW grant DIR/WK/2018/12.
      AAZ acknowledges support from the Polish National Science Center grants 2019/35/B/ST9/03944 and 2023/48/Q/ST9/00138, and from the Copernicus Academy grant CBMK/01/24. MB was funded in part by PRIN TEC INAF 2019 ``SpecTemPolar! -- Timing analysis in the era of high-throughput photon detectors’’.
      This research has made use of NASA’s Astrophysics Data System Bibliographic Service (ADS) and of ISIS functions (\texttt{isisscripts}\footnote{\url{http://www.sternwarte.uni-erlangen.de/isis/}}) provided by ECAP/Remeis observatory and MIT.
      The authors thank Mirjam Oertel, developer of the first versions of some of the scripts calculating the colour-colour tracks. They also thank John E. Davis for the development of the slxfig\footnote{http://www.jedsoft.org/fun/slxfig/} module used to prepare some of the figures in this work. Colour schemes used in the first part of this paper were based on Paul Tol’s color-blindness-friendly palettes and templates\footnote{https://personal.sron.nl/~pault/}. 
\end{acknowledgements}

%
%

\bibliographystyle{aa}
\bibliography{bib}

\begin{appendix}\section{}
\label{app_model}

The model used to describe the broad band primary continuum for simulations reported in Sects. \ref{complexcont} and \ref{stratified} includes a disc black body component, a soft Comptonisation component, and a hard Comptonisation component and its associated reflection, i.e. \texttt{TBabs} \texttt{$\times$ [diskbb + nthComp + relxillCp]} in \texttt{XSPEC}.
We froze the BH spin at the maximum value $a=0.998$ (which allows the inner disc truncation radius to span the largest range of values), the inclination of the reflector $i$ at $27^{\circ}$ (\citealt{MillerJones_2021}, see also recent results by \citealt[][]{Krawczynski_2022} and \citealt[][]{Poutanen_2023}) and the high-energy cut-off of the hard Comptonisation component at 100 keV \citep{Basak_2017}. 
The \texttt{nthComp} component is included in the model to account for the observed ``soft excess'', as done in e.g. \citet{Basak_2017}. 
The seed photon temperature of the soft Comptonisation component is tied to the best-fit inner disc temperature of the \texttt{diskbb} component.

\begin{figure}[!h]
\centering
\includegraphics[width=1.0\linewidth]{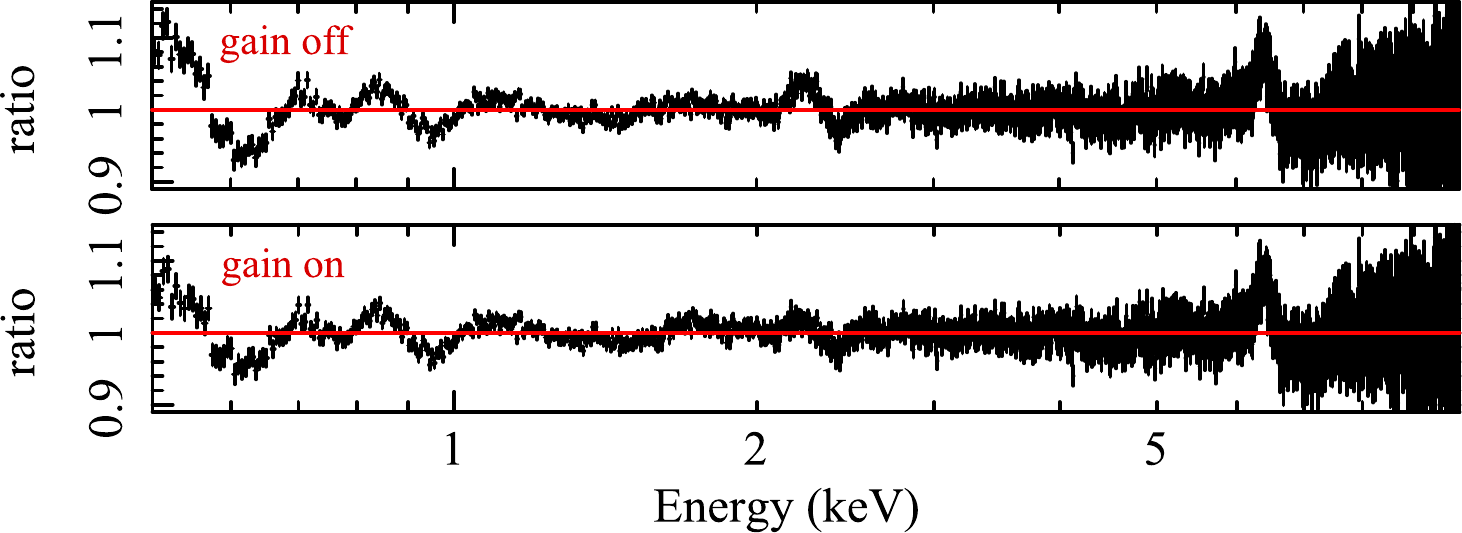}
    \caption{Comparison of the  data to model ratios before (upper panel) and after (bottom panel) using the \texttt{gain} correction (\texttt{intercept} parameter set to 0.01 keV and \texttt{slope} parameter fixed to 0).}
    \label{ratio_gain}
\end{figure}

This model was fit to the spectrum of the {\it XMM-Newton} observation 201 of Cyg~X-1 after filtering out strong dips (the NWA dataset selected in \citealt{Lai_2022}). 
After fitting this model, the data still show narrow residuals, mainly between ${\sim}0.5$--$2.5\,\rm{keV}$. Since these residuals appear close to the absorption edges of the response matrix, i.e. at $E \approx\,0.528$\,keV, and between $E \approx\,1.83$--$1.87\,\rm{keV}$, they are likely due to calibration problems.
In order to account for these residuals, we modified the response matrix using the \texttt{gain} function in \texttt{XSPEC}. This command permits to change the slope or the intercept of the effective area curve, shifting the energies at which the response matrix is defined. Therefore, we introduced a linear gain shift (\texttt{intercept} parameter in \texttt{XSPEC}) of 0.01 keV (\texttt{slope} parameter of \texttt{gain} fixed to 0). As a result, the best-fit strongly improves (from $\chi^2 / \mathrm{dof} = 4404.17/1891$ to $\chi^2 / \mathrm{dof} = 3923.7/1891$, see Fig.~\ref{ratio_gain}). 
\\
Nonetheless, some residuals were still present, including a strong narrow excess in the Fe K$\alpha$ region. 
To account for these features, we added narrow ($\sigma<0.1$ keV) emission and absorption gaussian components (see Tab.~\ref{tab_model}). We did not investigate the nature of these features, but we point out that they might be due to incorrect calibration or residual wind absorption after the selection process of the NWA GTIs \citep{Lai_2022}. On the other hand, the narrow feature observed in the Fe K$\alpha$ region might indicate the need to include a second reflection component (such as from the outer disc) to obtain a better description of the broad band continuum. Nonetheless, we stress that our analysis, which focuses on testing the effects of variable absorption on the colour-colour tracks, is not strongly influenced by the details of the continuum model as long as it provides a good description of the broad-band spectrum. 

\begin{figure}[!h]
\centering
\includegraphics[width=1.0\linewidth]{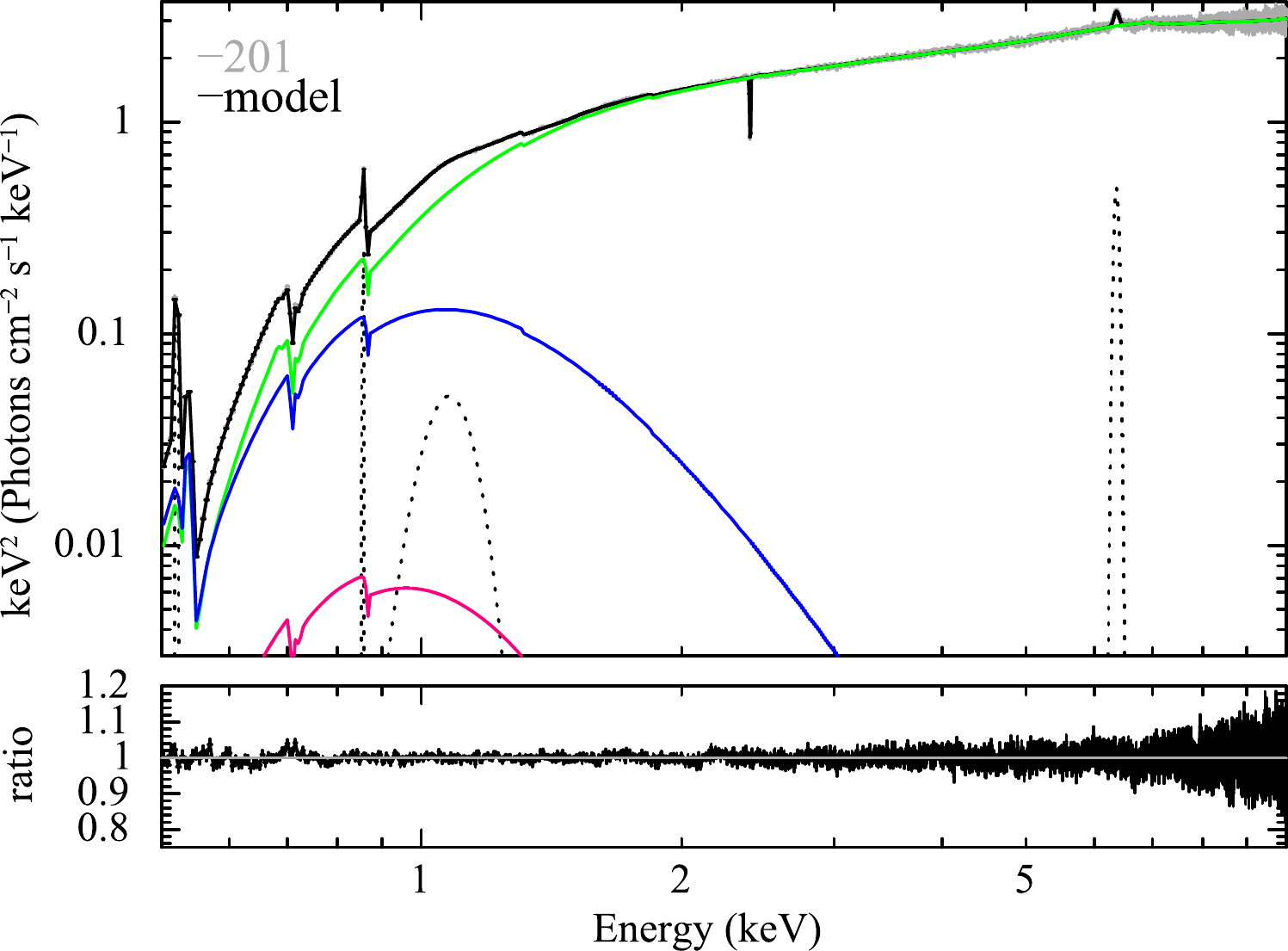}
    \caption{The best-fit model of the NWA time-averaged spectrum of observation 201. The complex best-fit model is overplotted in black, while the single components are shown in different colours: in magenta the disc blackbody, in blue the soft excess, in green the hard Comptonisation component and its relative reflection and in dotted grey the additional gaussians. Ratios of the data to the best-fitting model are shown in the bottom panel.}
    \label{spectrum}
\end{figure}

The final best-fit model yields $\chi^2 / \mathrm{dof} = 2263.76/1873$. 
Fig.~\ref{spectrum} shows the spectrum of the NWA time-averaged spectrum of observation 201 and its best-fit model. The most relevant parameters are listed in Tab.~\ref{tab_model}. 

\setlength{\tabcolsep}{0.9pt}
\renewcommand{\arraystretch}{1.5}
\begin{table}
	\centering
    \caption{Best-fit parameters of the continuum model, obtained from the fit of the NWA spectrum of observation 201. The errors are reported at the 90\% confidence level, i.e. $\Delta\chi^2=2.71$.} 
	\label{tab_model}

	\begin{tabular}{cccc}
		\hline
		& Component & Parameter & Value\\
		   
		\hline
		ISM absorption & \texttt{TBabs}  & $N_{\rm H}$ $(10^{22} \rm{cm}^{-2})$ & $0.79_{-0.02}^{+0.01}$\\
        \hline
		\texttt{continuum} & \texttt{diskbb} & $kT_{bb}$ (keV) & $<0.14$ \\
	                       & \texttt{nthComp} & $\Gamma$ &  $<5.9$ \\      
	                       & & $kT_{e}$ (keV) & $>0.67$ \\  
			               & \texttt{relxillCp} & $R_{in}$ ($\rm{R_g}$)  & $ 13.24\pm1.26$\\ 
			                                    & & $\Gamma$ & $1.59\pm0.01$\\
		                                        & & $\log_{10}\xi$  & $3.32\pm0.01$ \\
		                                        & & $A_{Fe}$ & $1.01_{-0.02}^{+0.05}$\\
		                                        & & $\mathcal{R}$ & $0.52_{-0.03}^{+0.01}$ \\
		                                    
		\hline

    	\footnotesize{\texttt{absorption features}}
	                      & \texttt{gau1} &  $E_l$ (keV) &  $1.44\pm0.01$\\
		                              
                          & \texttt{gau2} &  $E_l$ (keV) &  $2.40\pm0.01$\\
		                              
		\hline                             
		  				
		\footnotesize{\texttt{emission features}} & \texttt{gau3} &  $E_l$ (keV) & $6.36\pm0.01$ \\
		                             
		                & \texttt{gau4} & $E_l$ (keV) &  $0.52\pm0.01$\\
		                              
                        & \texttt{gau5} & $E_l$ (keV) &  $0.86\pm0.01$\\
		                              
		                & \texttt{gau6} &  $E_l$ (keV) & $1.05\pm0.01$\\

		\hline                        
	\end{tabular}

\end{table}

\section{Time-resolved colour-colour diagrams: best-fit models}

\begin{figure*}
\centering 
\includegraphics[width=1\linewidth]{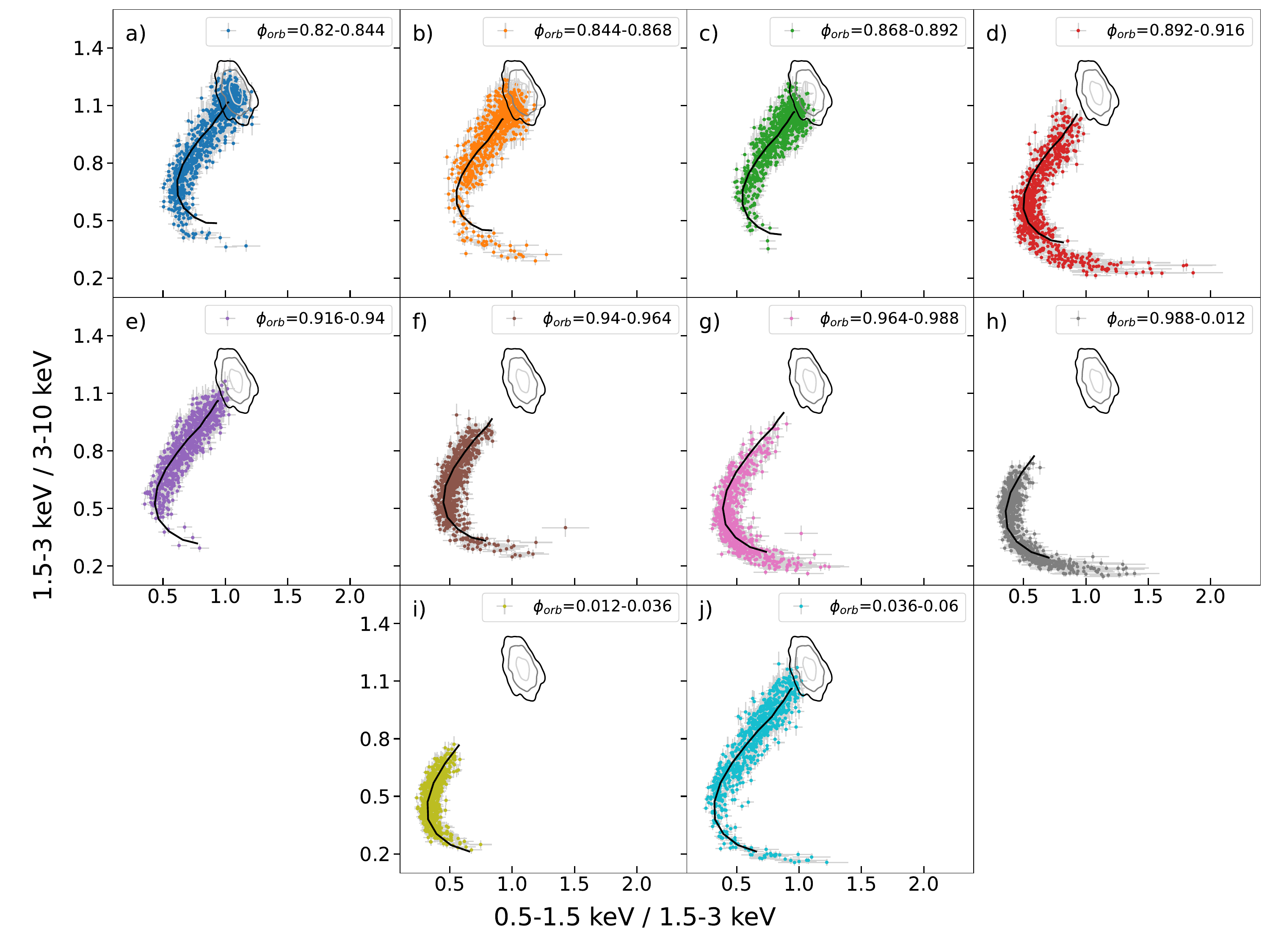}
    \caption{Time-resolved colour-colour diagrams of observation 201. Overplotted (black tracks) are the best-fit models obtained assuming $\log\xi=\log\frac{10+[N_{\rm{H,w}}/10^{22}\,\rm{cm}^{-2}]}{[N_{\rm{H,w}}/10^{22}\,\rm{cm}^{-2}]}+1$.
    The contour plots represent the 99.7\% (in black), 95\% (in grey) and 68\% (in light grey) confidence levels of the data distribution during the least absorbed stages of the orbit ($\phi_{\mathrm{orb}}=0.43-0.46$, see Appendix \ref{app_501}).} 
    \label{time_resolved_simulated_tracks}
\end{figure*}

In Fig.~\ref{time_resolved_simulated_tracks} we show the time-resolved colour-colour diagrams of observation 201 (colour-coded as in Fig.~\ref{time_resolved_CCDiagram}), and the best-fit models obtained from the fits discussed in Sect.~\ref{time-resolved}.

\section{Probability distribution close to inferior conjunction}
\label{app_501}
\begin{figure}[!h]
\centering 
\includegraphics[width=1\linewidth]{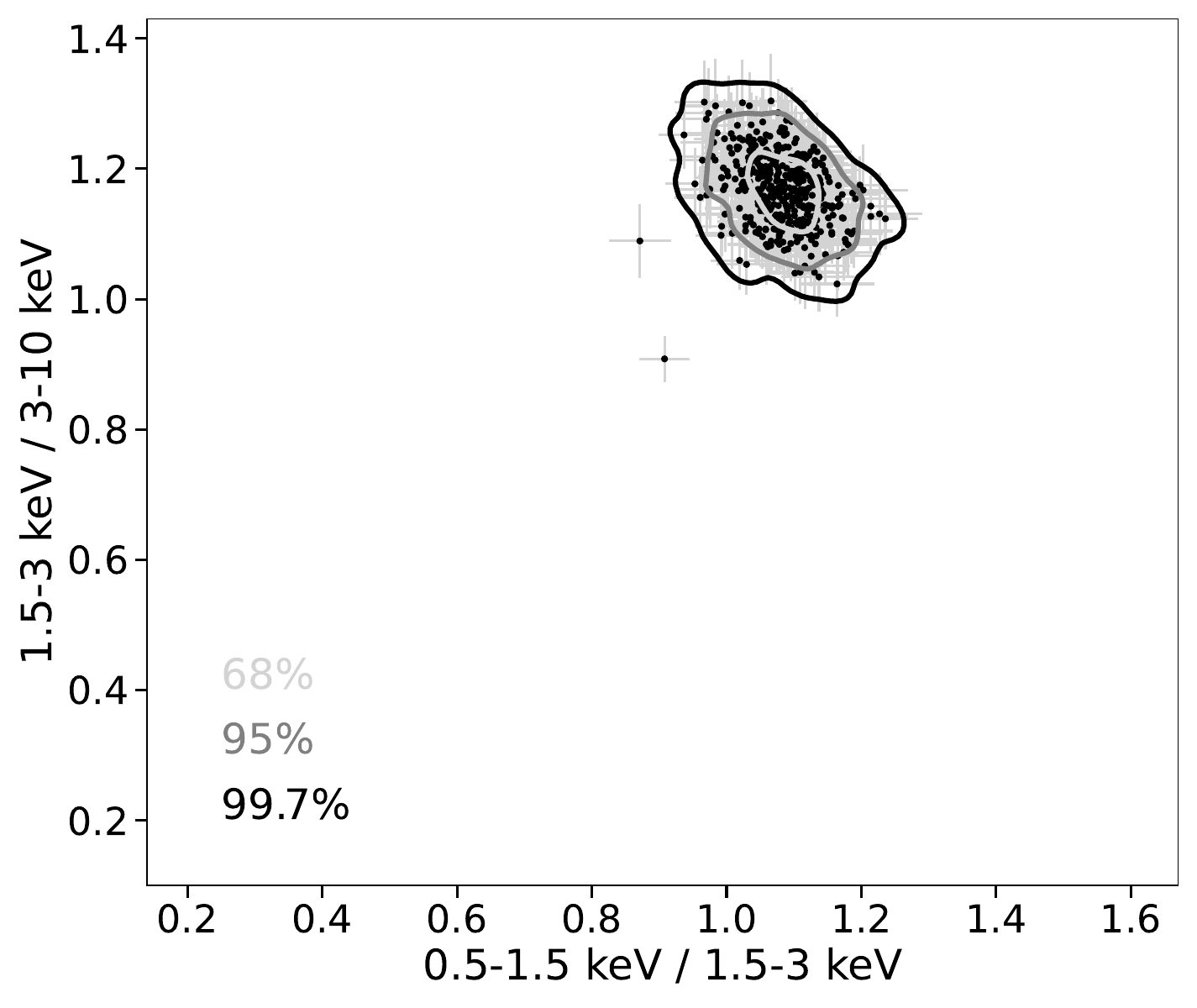}
     \caption{Colour-colour diagram and probability distribution map for orbital phases $\phi_{\mathrm{orb}}=0.43$--$0.46$. The curves correspond to the 99.7\% (in black), 95\% (in grey) and 68\% (in light grey) confidence regions.} 
    \label{kde_501_area}
\end{figure}

We show the colour-colour diagram calculated from orbital phases $\phi_{\mathrm{orb}}=0.43$--$0.46$, covered by observation 501 \citep{Lai_2022}. These phases are selected to be the closest to inferior conjunction (i.e. $\phi_{\mathrm{orb}} \sim 0.5$), thus the least affected by wind absorption.
Using the KDE method, we computed the probability distribution for this dataset (Sect.~\ref{time-resolved}). Fig.~\ref{kde_501_area} shows the resulting 99.7\%, 95\% and 68\% confidence contours overplotted to the data. 

\end{appendix}

\end{document}